\documentclass[aps,prd,showpacs]{revtex4}

\usepackage{amsmath,amsfonts,amssymb}
\usepackage[dvips]{graphicx}

\newtheorem{definition}{Definition}

\newcommand{\ud}{\mathrm{d}}

\begin{document}

\title{Cosmological zoo -- accelerating models with dark energy} 
\author{Marek Szyd{\l}owski}
\email{uoszydlo@cyf-kr.edu.pl}
\affiliation{Astronomical Observatory, Jagiellonian University, Orla 171, 
30-244 Krak{\'o}w, Poland}
\affiliation{Marc Kac Complex Systems Research Center, Jagiellonian University, 
Reymonta 4, 30-059 Krak{\'o}w, Poland}

\begin{abstract}
Recent observations of type Ia supernovae indicate that the Universe is in
an accelerating phase of expansion. The fundamental quest in theoretical 
cosmology is to identify the origin of this phenomenon. In principle there are 
two possibilities: 1) the presence of matter which violates the strong energy
condition (a substantial form of dark energy), 2) modified Friedmann equations 
(Cardassian models -- a non-substantial form of dark matter). We classify all 
these models in terms of $2$-dimensional dynamical systems of the Newtonian
type. We search for generic properties of the models. It is achieved with
the help of Peixoto's theorem for dynamical system on the Poincar{\'e} sphere.
We find that the notion of structural stability can be useful to
distinguish the generic cases of evolutional paths with acceleration. We find
that, while the $\Lambda$CDM models and phantom models are typical accelerating
models, the cosmological models with bouncing phase are non-generic in the
space of all planar dynamical systems. We derive the universal shape of
potential function which gives rise to presently accelerating models. Our
results show explicitly the advantages of using a potential function (instead
of the equation of state) to probe the origin of the present acceleration. We
argue that simplicity and genericity are the best guide in understanding our
Universe and its acceleration.
\end{abstract}


\pacs{98.80.Bp,98.80.Cq}

\maketitle

\section{Introduction}

As indicated by observations of distant type Ia supernovae
\cite{Riess:1998cb,Perlmutter:1998np} and other complementary observations 
\cite{Spergel:2003cb,Tegmark:2003ud} our Universe is presently accelerating. 
The problem of accelerated expansion of the current universe seems to be one of 
the most fundamental problems of theoretical physics of the XXI century (for 
a recent review see \cite{Padmanabhan:2002ji,Copeland:2006wr} and references 
therein).

There are several propositions of explanation of the observational fact that 
the Universe is speeding up rather than slowing down. They involve the 
cosmological constant, time dependent vacuum energy, dynamical scalar field 
(quintessence) or modified Friedmann equations
\cite{Freese:2002sq,Godlowski:2003pd,Nojiri:2006ri}. However, in all truth, we 
do not know what is causing the effect. While the proposition of the 
cosmological constant is an attractive idea from both the theoretical and 
observational point of view, it entails a crucial problem. Why is its value, 
as measured by distant SNIa observations, so small when compared to the value 
obtained from the quantum field theory considerations? Neither do we understand 
why the dark energy density is of the same order of magnitude as the matter 
density during the present epoch.

Recently some new developments in dark energy studies were proposed. For 
example non-Abelian Einstein-Born-Infeld dilaton theory was explored 
\cite{Fuzfa:2005qn,Fuzfa:2006pn}, as well as cosmology based on its 
generalization \cite{Troisi:2006ur}.

In this work, we start form general relativity cosmological models with 
Robertson--Walker symmetry rather than Euler's and Poison's equations. We adopt 
the particle--like description of FRW cosmologies \cite{Lima:1998kw} which we 
extend to a general class of FRW cosmologies with dark energy component
parameterized by the scale factor (or redshift). The main advantage of such
formulation is that the one dimensional potential function contains the
complete information about the dynamics. Finally, we obtain a unified
description of a large class of cosmological models in the notion of a particle
moving in a one-dimensional potential well. Hence, the evolution of the system
can be naturally reduced to a simple $2$-dimensional dynamical system of the 
Newtonian type.

The problem of structural stability was studied by Lazkoz in the context of 
cosmic acceleration \cite{Aguirregabiria:2004eu,Lazkoz:2004rc}. The ideas of 
rigidity and  fragility of solutions are translated from the language of 
structural stability which was introduced by Andronov and Pontryagin 
\cite{Andronov:1937sg}. Following Lidsey \cite{Lidsey:1992wk} the concepts 
of rigidity and fragility should be described through a condition on the 
functional form of the Hubble function. In this paper we formulate the condition 
for structural stability in terms of the potential function rather than 
the Hubble function. However, the simple relation between them makes the 
equivalence of this approaches.

All attempts at cosmological modelling involve at the very beginning a large
number of simplification and theoretical assumption regarding the parameters of
the dark energy of unknown form. The question then arises whether such simple
models are representing properties of real systems. Usually, for dynamical
systems to be viable as models they need to be structurally stable, i.e. their 
dynamics must preserve its qualitative characteristics under small perturbation 
\cite{Tavakol:1988,Tavakol:1990st}. Such a point of view resolves the 
``approximation problem'' because our models of reality are, by definition, not 
precise, but might be very close to it. Moreover, the structural stability 
framework can be useful from the methodological point of view because 
\textit{a'priori\/} we do not know the functional forms of dynamical systems. 
Of course there is no objective reason to believe that all dynamical systems 
should be structurally stable. From the philosophical view of point, it is a 
presumption of convenience (or economy). According to Occam's Razor, simple 
theories are more economical and mostly cases a little improvement of 
prediction is paid a very large increase in complexity of a model. To 
discriminate between different dark energy models, we postulate a simplicity 
principle that among all dynamical laws describing the cosmological evolution, 
the laws with the smallest complexity are chosen. It is interesting that 
Occam's Razor became a cornerstone of modern theory of induction 
\cite{Solomonoff:1964a,Solomonoff:1964b}. The notion of structural stability 
was discussed in the context of Kaluza-Klein theories 
\cite{Deruelle:1986ia,Deruelle:1987ux}. In this class of models it is required 
the existence of the mechanism of dynamical reduction of extra-dimensions, 
i.e. the configuration of FRW $\times$ \{static internal space\/\} should be an 
attractor. 

In this paper we apply the structural stability notion as the discriminatory 
among the cosmological models with acceleration phase of expansion. In other 
words structural stability acting as Occam's Razor allows us to choose the 
simplest (the $\Lambda$CDM or phantom CDM) cosmological models with 
acceleration. From a physical view of point these two are two-phased models 
with decelerating matter-dominated phase and accelerating dark energy dominated 
phase. 

It is interesting that the strong energy condition $(\rho + 3p) \ge 0$ and 
$(\rho + p) \ge 0$ (SEC) which determines qualitative dynamics of the FRW 
cosmological models can be translated into the constraint on the potential 
function \cite{Szydlowski:2005uq}
\[
I_{V}(a) = \frac{d\ln V}{d\ln a} \ge 0 \quad \text{and} \quad I_{V}(a) \le 2
\]
where $I_{V}(a)$ is an elasticity coefficient which measures the logarithmic 
slope of the potential function with respect to the scale factor $a$. 

The famous Hawking-Penrose singularity theorems invoke this condition 
\cite{Hawking:1973}. Note that the null energy condition (NEC) $\rho + p \ge 0$ 
involving a derivative of the potential function can be formulated in an exact 
form 
\[
-\frac{dV}{da} + 2\frac{V(a)}{a} > 0, \quad V(a) \le \lambda a^2, \quad 
\lambda = \text{const} > 0.
\]
The violation of the SEC means that $V(a)$ is a decreasing function of the 
scale factor. The violation of the NEC denotes that in some region the diagram 
of the potential function is over the diagram of a parabola $V(a) \propto a^2$. 

The idea of obtaining a Newtonian analogy to the FRW cosmology within the
framework of classical mechanics has been considered since the classical
papers of Milne and McCrea \cite{Milne:1934,McCrea:1951}. In their
formulation of the cosmological problem the fluid dynamics for a dust filled
universe is applied but there arises a crucial difficulty because pressure $p$
does not play a dynamical role like in general relativity theory. As a
consequence, there is no classical analogy of a radiation filled universe.

If we assume the validity of the Robertson-Walker symmetry for our universe 
which is filled with perfect fluid satisfying the general form of the equation 
of state $p_{\text{eff}}=w_{\text{eff}}(a)\rho_{\text{eff}}$, then
$\rho_{\text{eff}}=\rho_{\text{eff}}(a)$ i.e. both the effective energy
density $\rho_{\text{eff}}$ and pressure $p_{\text{eff}}$ are parameterized
by the scale factor as a consequence of the conservation condition
\begin{equation}
\dot{\rho}_{\text{eff}}=-3 H (\rho_{\text{eff}} + p_{\text{eff}}),
\label{eq:1}
\end{equation}
where dot denotes differentiation with respect to the cosmological time $t$
and $H=(\ln a)\dot{}$ is the Hubble function.

To unify two main approaches to explaining SNIa data, i.e. 1) dark energy,
and 2) modification of FRW equations, we generalize the acceleration equation
\begin{equation}
\frac{\ddot{a}}{a}=-\frac{1}{6}(\rho_{\text{eff}}+3p_{\text{eff}}) + 
\frac{B}{6} a^{m},
\label{eq:2}
\end{equation}
where $B$ and $m$ are constants, rather than doing so for the first Friedmann
equation like in Freese and Lewis's approach.

We assume that the Universe is filled with standard dust matter (together with
dark matter) and dark energy $X$
\begin{equation}
\begin{array}{l}
p_{\text{eff}} = 0 + w_{X}\rho_{X},\\
\rho_{\text{eff}} = \rho_{\text{m}} + \rho_{X},
\end{array}
\label{eq:3}
\end{equation}
where $w_{X}=w_{X}(a)$ is the coefficient of the equation of state for dark 
energy parameterized by the scale factor or redshift $z\colon 1+z=a^{-1}$.

One can check that the Raychaudhuri equation (\ref{eq:2}) can be rewritten to
the form analogous to the Newtonian equation
\begin{equation}
\ddot{a}=-\frac{\partial V}{\partial a},
\label{eq:4}
\end{equation}
if we choose the following form of the potential function $V(a)$
\begin{equation}
\left\{ \begin{array}{lll}
\displaystyle{-\frac{1}{6}(\rho_{\text{eff}} a^{2} + \frac{B}{m+2} a^{m+2})} 
& \text{for} & m \ne -2, \\
\displaystyle{-\frac{1}{6}(\rho_{\text{eff}} a^{2} + B \ln{a})} 
& \text{for} & m=-2,
\end{array} \right.
\label{eq:5}
\end{equation}
where $\rho_{\text{eff}}$ satisfies the conservation condition (\ref{eq:1}). 
As the alternative method to obtain (\ref{eq:5}) is integration by parts 
eq. (\ref{eq:4}) with the help of conservation condition (\ref{eq:1}).

If we put $B=0$ in formula (\ref{eq:5}) then we obtain the standard cosmology
with dark energy. If we consider $m \ne -2$,
$\rho_{\text{eff}}=\rho_{\text{m}}$, then the Cardassian cosmology can be
recovered provided that $\rho=\rho(a)$ is assumed. The case of $B \ne 0$ and
$m=-2$ represents a new exceptional case which appears as a consequence of
the generalized Raychaudhuri equation instead of the Friedmann first integral
which assumes the following form
\begin{equation}
\rho_{\text{eff}} - 3 \frac{\dot{a}^{2}}{a^{2}} = 3 \frac{k}{a^{2}} 
- \frac{B}{m+2} a^{m},
\label{eq:6}
\end{equation}
or
\begin{equation}
\dot{a}^{2} = -2 V
\label{eq:7}
\end{equation}
where
\begin{equation}
V = - \frac{\rho_{\text{eff}} a^{2}}{6} + \frac{k}{2} 
- \frac{B a^{m+2}}{6 (m+2)}, \quad m \ne -2.
\label{eq:8}
\end{equation}
Equation~(\ref{eq:7}) is the form of the first integral of the Einstein
equation with Robertson-Walker symmetry called the Friedmann energy first 
integral. Formally, curvature effects as well as the Cardassian term can be
incorporated into the effective energy density
$(\rho_{k}=-\frac{k}{a^{2}}, \rho_{\text{Card}}=\frac{B}{m+2} a^{m})$.

The form of equation (\ref{eq:4}) suggests a possible interpretation of the
evolutional paths of cosmological models as the motion of a fictitious particle
of unit mass in a one dimensional potential parameterized by the scale factor.
Following this interpretation the Universe is accelerating in the domain of
configuration space $\{a \colon a \ge 0\}$ in which the potential is a decreasing
function of the scale factor. In the opposite case, if the potential is an
increasing function of $a$, the Universe is decelerating. The limit case of
zero acceleration corresponds to an extremum of the potential function.
The energy conditions were also confronted with SNIa observations 
\cite{Santos:2006ja}. It is shown that all energy conditions seem to have 
been violated in a recent past of evolution of the Universe. 

It is useful to represent the evolution of the system in terms of dimensionless
density parameters $\Omega_{i} \equiv \rho_{i}/(3 H_{0}^{2})$, where $H_{0}$ is
the present value of the Hubble function. For this aims it is sufficient to
introduce the dimensionless scale factor $x \equiv a/a_{0}$ which measures the
value of $a$ in the units of the present value $a_{0}$, and parameterize the 
cosmological time following the rule $t \mapsto \tau \colon \ud t |H_{0}| = 
\ud \tau$. Hence we obtain a $2$-dimensional dynamical system describing the 
evolution of cosmological models
\begin{subequations}
\label{eq:9}
\begin{align}
\frac{\ud x}{\ud \tau} &= y,\\
\frac{\ud y}{\ud \tau} &= -\frac{\partial V}{\partial x},
\end{align}
\end{subequations}
and $y^{2}/2 + V(x)=0$, $1+z=x^{-1}$. Where
\[
V(x) = -\frac{1}{2}\left\{ \Omega_{\text{eff}} x^{2} 
+ \Omega_{\text{Card},0} x^{m+2} + \Omega_{k,0}\right\}, 
\]
\[
\Omega_{\text{eff}} = \Omega_{\text{m},0} x^{-3} + \Omega_{X,0}
x^{-3(1+w_{X})},
\] 
for dust matter and quintessence matter satisfying the equation
of state $p_{X}= w_{X} \rho_{X}$, $w_{X}=\text{const}$.

The form (\ref{eq:9}) of a dynamical system opens the possibility of adopting
the dynamical systems methods in investigations of all possible evolutional
scenarios for all possible initial conditions. Theoretical research in this
area has obviously shifted from finding and analyzing particular cosmological
solution to investigating a space of all admissible solutions and discovering
how certain properties (like, for example, acceleration, existence of
singularities) are ``distributed'' in this space.

The system (\ref{eq:9}) is a Hamiltonian one and adopting the Hamiltonian 
formalism to the admissible motion analysis seems to be natural. The analysis 
can then be performed in a manner similar to that of classical mechanics. The 
cosmology determines uniquely the form of the potential function $V(x)$, which 
is the central point of the investigations. Different potential functions for 
different propositions of solving the acceleration problem are presented in 
Table \ref{tab:1} and Table \ref{tab:2}.

\begin{table}
\caption{The potential functions for different dark energy models.}
\begin{ruledtabular}
\begin{tabular}{|c|c|c|}
\hline
 The model & The form of the potential function & Independent parameters \\
\hline \hline
Einstein-de Sitter model & & \\
$\Omega_{\text{m},0}=1$, $\Omega_{\Lambda,0}=0$ & $V(x)=
-\frac{1}{2}\Omega_{\text{m},0}x^{-1}$ & $H_{0}$\\
$\Omega_{k,0}=0$ & & \\
\hline
$\Lambda$CDM model & & \\
$\Omega_{\text{m},0}+\Omega_{\Lambda,0}=1$ & $V(x)=
-\frac{1}{2}\bigg\{\Omega_{\text{m},0}x^{-1} 
+ \Omega_{\Lambda,0}x^{2}\bigg\}$ & $(\Omega_{\text{m},0},H_{0})$\\
 & & \\
\hline
FRW model filled with $n$ & & \\
noninteracting multi-fluids & $V(x)=-\frac{1}{2}\bigg\{\Omega_{\text{m},0}x^{-1} + \Omega_{k,0} + \sum_{i=1}^{N} \Omega_{i,0} x^{-3(1+w_{i})}\bigg\}$ & $(\Omega_{\text{m},0},\Omega_{\Lambda,0},\Omega_{n,0},H_{0})$\\
$p=w\rho$ with dust matter & & \\
and curvature & & \\
\hline
FRW quintessence model with & & \\
dust and dark matter $X$ & $V(x)=-\frac{1}{2}\bigg\{\Omega_{\text{m},0}x^{-1}+\Omega_{k,0}+\Omega_{X,0}x^{-1-3w_{X}}\bigg\}$ & $(\Omega_{\text{m},0},\Omega_{X,0},H_{0})$ \\
$w_{X}<-1$ phantom models & & $\Omega_{k,0}=1-\Omega_{\text{m},0}-\Omega_{X,0}$ \\
\hline
FRW model with generalized & & \\
Chaplygin gas \cite{Kamenshchik:2001cp,Biesiada:2004td} & $V(x)=-\frac{1}{2}\bigg\{\Omega_{\text{m},0}x^{-1}+\Omega_{k,0}+\Omega_{\text{Chapl},0}(A_{S} + \frac{1 - A_{S}}{x^{3(1+\alpha)}})^{\frac{1}{1+\alpha}} \bigg\}$ & $(\Omega_{\text{m},0},\Omega_{\text{Chapl},0},H_{0})$ \\
$p=-\frac{A}{\rho^{\alpha}}$, $A>0$ & & \\
\hline
FRW models with dynamical & & \\
equation of state for dark energy & $V(x)=-\frac{1}{2}\bigg\{\Omega_{\text{m},0}x^{-1} + \Omega_{k,0} + \Omega_{X,0}x^{-1}\exp{[\int_{1}^{x}\frac{w_{X}(a)}{a}\ud a]} \bigg\}$ & $(\Omega_{\text{m},0},\Omega_{X,0},H_{0})$ \\
$p_{X}=w_{X}(a)\rho_{X}$ and dust & & \\
\hline
FRW models with dynamical & & \\
equation of state for dark energy & $V(z)=-\frac{1}{2}\left\{\Omega_{\text{m},0}(1+z)+\Omega_{X,0}(1+z)^{1+3(w_{0}-w_{1})} e^{3w_{1}z} +\Omega_{k,0} \right\}$ & $(\Omega_{X,0},w_{0},w_{1},H_{0})$\\
coefficient equation of state & & \\
$w_{X}=w_{0}+w_{1}z$ & & \\
\hline
\end{tabular}
\end{ruledtabular}
\label{tab:1}
\end{table}

\begin{table}
\caption{The potential functions for different cosmological models which offer the possibility of explanation of acceleration in terms of modified FRW equations.}
\begin{ruledtabular}
\begin{tabular}{|c|c|c|}
\hline
 The model & The form of the potential function & Independent parameters \\
\hline \hline
Non-flat Cardassian models & & \\
filled by dust matter \cite{Freese:2002sq,Godlowski:2003pd}& 
$V(x)=-\frac{1}{2}\left\{\Omega_{\text{m},0}x^{-1} + \Omega_{k,0}+\Omega_{\text{Card},0} x^{m+2}\right\}$ & 
$(H_{0},\Omega_{\text{m},0},\Omega_{\text{Card},0})$ \\
\hline
Bouncing cosmological models & & \\
$(H/H_{0})^{2}=\Omega_{\text{m},0}x^{-m}-\Omega_{n,0}x^{-n}$, & $V(x)=-\frac{1}{2}\left\{\Omega_{\text{m},0}x^{-m+2}-\Omega_{n,0}x^{-n+2} \right\}$ & $(H_{0},\Omega_{\text{m},0},\Omega_{n,0},m,n)$ \\
$n>m$ \cite{Szydlowski:2005qb} & & \\
\hline
Randall-Sundrum brane & & \\
models with dust on the brane & $V(x)=-\frac{1}{2}\left\{\Omega_{\text{m},0}x^{-1}+\Omega_{\lambda,0}x^{-6}+\Omega_{k,0}+ \Omega_{d,0}x^{-4} \right\}$ & $(H_{0},\Omega_{\text{m},0},\Omega_{\lambda,0})$ \\
and dark radiation \cite{Randall:1999vf} & & \\
\hline
Cosmology with spin & & \\
and dust & 
$V(x)=-\frac{1}{2}\left\{\Omega_{\text{m},0}x^{-1} + \Omega_{s,0}x^{-6} + \Omega_{k,0} \right\}$ & 
$(H_{0},\Omega_{s,0},\Omega_{k,0})$ \\
(MAG cosmology) \cite{Krawiec:2005jj} & & \\
\hline
Dvali, Deffayet, Gabadadze & & \\
brane models & $V(x)=-\frac{1}{2} \left\{\sqrt{\Omega_{\text{m},0}x^{-1}+\Omega_{rc,0}} + \sqrt{\Omega_{rc,0}}\right\}^{2}$ & $(H_{0},\Omega_{\text{m},0})$\\
(DDG) \cite{Deffayet:2001pu} & & \\
\hline
Sahni, Shtanov & & \\
brane models \cite{Shtanov:2000vr,Sahni:2002dx} & $V(x)=-\frac{1}{2}\bigg\{\Omega_{\text{m},0}x^{-1} + \Omega_{\sigma,0} + 2\Omega_{l,0}$ & $(H_{0},\Omega_{\text{m},0},\Omega_{\sigma,0},\Omega_{l,0},\Omega_{\Lambda_{b},0})$ \\
 & $\pm 2\sqrt{\Omega_{l,0}}\sqrt{\Omega_{\text{m},0}x^{-1}+\Omega_{\sigma,0}+\Omega_{l,0}+\Omega_{\Lambda_{b},0}} \bigg\}$ & \\
\hline
FRW cosmological models & & \\
of nonlinear gravity $\mathcal{L}\propto R^{n}$ & $V(x)=-\frac{1}{2}
\left\{\frac{2n}{3-n}\Omega_{\text{m},0}x^{-1}
+\frac{4n(2-n)}{(n-3)^{2}}\Omega_{r,0}x^{-2} \right\} 
\Omega_{\text{nonl},0}x^{\frac{3(n-1)}{n}}$ & 
$(H_{0}, \Omega_{\text{m},0}, \Omega_{\text{nonl},0})$ \\
with matter and radiation \cite{Allemandi:2004ca,Capozziello:2004vh} & & \\
\hline
$\Lambda$DGP model \cite{Deffayet:2000uy,Lue:2004za} & & \\
screened cosmological & $V(x)=-\frac{1}{8}x^{2}\left\{-\frac{1}{r_{0}H_{0}}
+\sqrt{(2+\frac{1}{r_{0}H_{0}})^{2}+4\Omega_{\text{m},0}(x^{-3}-1)} \right\}$ 
& $(r_{0}H_{0},H_{0})$\\
constant model & & \\
\hline
\end{tabular}
\end{ruledtabular}
\label{tab:2}
\end{table}
 
The key problem of this paper is to investigate the geometrical and topological
properties of the multiverse of models of accelerating universes which we 
define in the following way
\begin{definition}
By multiverse of models of accelerating Universes we understand the space of 
all $2$-dimensional systems of the Newtonian type $\dot{x}=y$, $\dot{y}=
-\partial V/\partial x$ with suitably defined potential function of the scale 
factor, which characterize the physical model of the dark energy or 
modification of the FRW equation.
\end{definition}

The organization of the text is the following. In section~\ref{sec:2} we
investigate the property of structural stability of different subsets of
multiverse. Section~\ref{sec:3} is devoted to investigation of the inverse
problem of accelerating cosmology -- reconstruction of potential function from
SNIa data and estimation of the value of transition redshift and Hubble
function at the moment of changing decelerating phase into accelerating one. In
section~\ref{sec:4} we distinguish generic accelerating models of the
multiverse with the help of Peixoto's theorem which characterize the 
structurally table systems on the compact two-dimensional phase plane defined 
on a two-dimensional Poincar{\'e} sphere. In section~\ref{sec:5} we formulate 
the conclusions and examine the significance of the obtained results for the 
philosophical discussion over McMullin's indifference principle, and 
the fine-tuning principle in the modern cosmological context.

\section{Structural stability issues}
\label{sec:2}

Einstein's field equations constitute, in general, a very complicated system of
nonlinear, partial differential equations, but what is made use of in cosmology
are the solutions with prior symmetry assumptions postulated at the very
beginning. In this case, the  Einstein field equations can be reduced to a 
system of ordinary differential equation, i.e. a dynamical system. Hence, in
cosmology the dynamical systems methods can be applied in a natural way. The
applications of these methods allow to reveal some stability properties of
particular solutions, visualized geometrically as trajectories in the
phase space. Hence, one can see how large the class of the solutions leading
to the desired property is, by means of attractors and the inset of limit set 
(an attractor is a limit set with an open inset -- all the initial conditions 
that end up in some equilibrium state). The attractors are the most prominent
experimentally, because of the probability for an initial state of the
experiment to evolve asymptotically to the limit set being proportional to the
volume of the inset.

The idea, now called structural stability, emerged in the history of
dynamical investigations in the 1930's with the writings of Andronov,
Leontovich and Pontryagin in Russia (the authors do not use the name
structural stability but rather the name ``roughly systems''). This idea is
based on the observation that actual state of the system can never be specified
exactly and application of dynamical systems might be useful anyway if it can
describe features of the phase portrait that persist when state of the system
is allowed to move around (see Ref.~\cite[p. 363]{Abraham:1992dg} for more
comments).

Among all dynamicists there is a shared prejudice that
\begin{enumerate}
\item there is a class of phase portraits that are far simpler than arbitrary
ones which can explain why a considerable portion of the mathematical physics
has been dominated by the search for the generic properties. The exceptional
cases should not arise very often in application and they de facto interrupt
discussion (classification) \cite[p. 349]{Abraham:1992dg};
\item the physically realistic models of the world should posses some kind of
structural stability because to have many dramatically different models all
agreeing with observations would be fatal for the empirical methods of science
\cite{Thom:1977ss,Szydlowski:1984ss,Biesiada:2003tp,Golda:1987ip,Tavakol:1988,
Farina-Busto:1990kh,Tavakol:1991,Coley:1992}.
\end{enumerate}

In cosmology a property (for example acceleration) is believed to be physically
realistic if it can be attributed to generic subset of the models within a
space of all admissible solutions or if it possesses a certain stability, i.e.
if it is shared by a ``epsilon perturbed model''. For example G.~F.~R. Ellis
formulates the so called probability principle ``The universe model should be 
the one that is a probable model'' within the set of all universe models and 
a stability assumption which states that ``the universe should be stable to 
perturbations'' \cite{Szydlowski:1984ss}.

The problem is how to define
\begin{enumerate}
\item a space of states and their equivalence,
\item a perturbation of the system.
\end{enumerate}

The dynamical system is called structurally stable if all its 
$\delta$-perturbations (sufficiently small) have an epsilon equivalent phase
portrait. Therefore for the conception of structural stability we consider a
$\delta$ perturbation of the vector field determined by right hand sides of 
the system which is small (measured by delta). We also need a conception of 
epsilon equivalence. This has the form of topological equivalence -- a 
homeomorphism of the state space preserving the arrow of time on each 
trajectory. In the definition of structural stability consider only 
deformation of the ``rubber sheet'' type stretches or slides of the phase 
space by a small amount measured by epsilon.

\begin{figure}
a)\includegraphics[scale=0.5,angle=0]{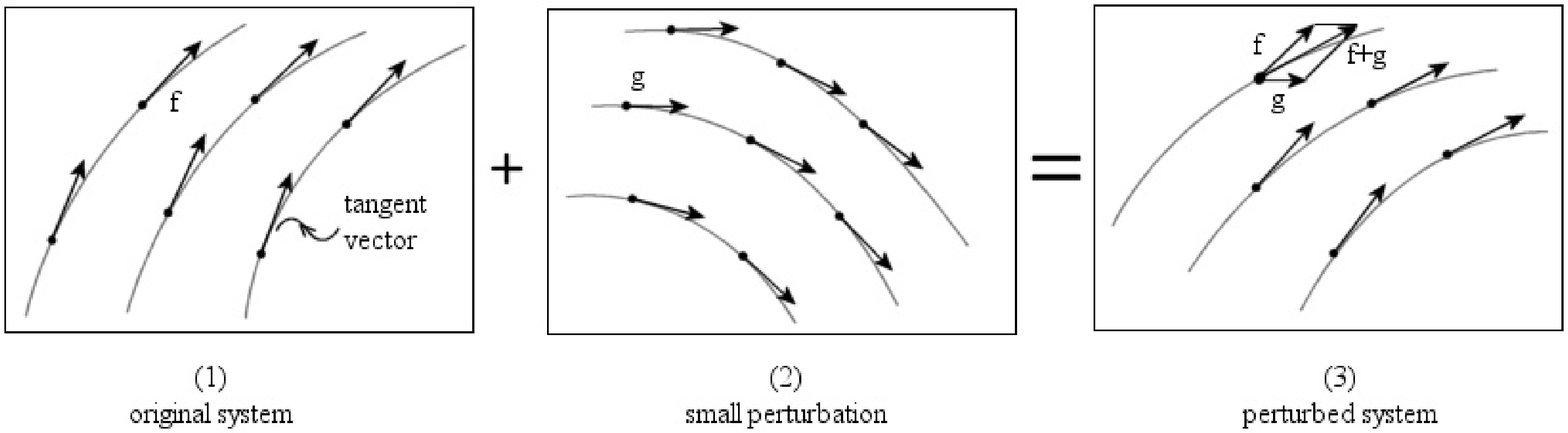}
b)\includegraphics[scale=1,angle=0]{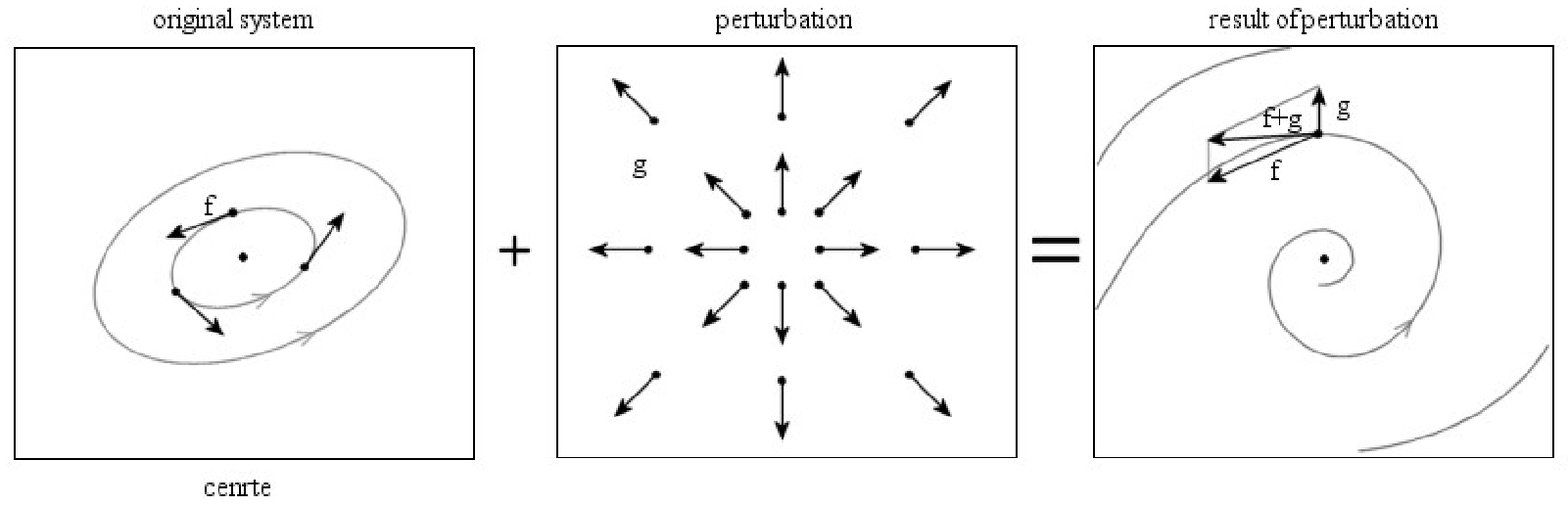}
c)\includegraphics[scale=0.5,angle=0]{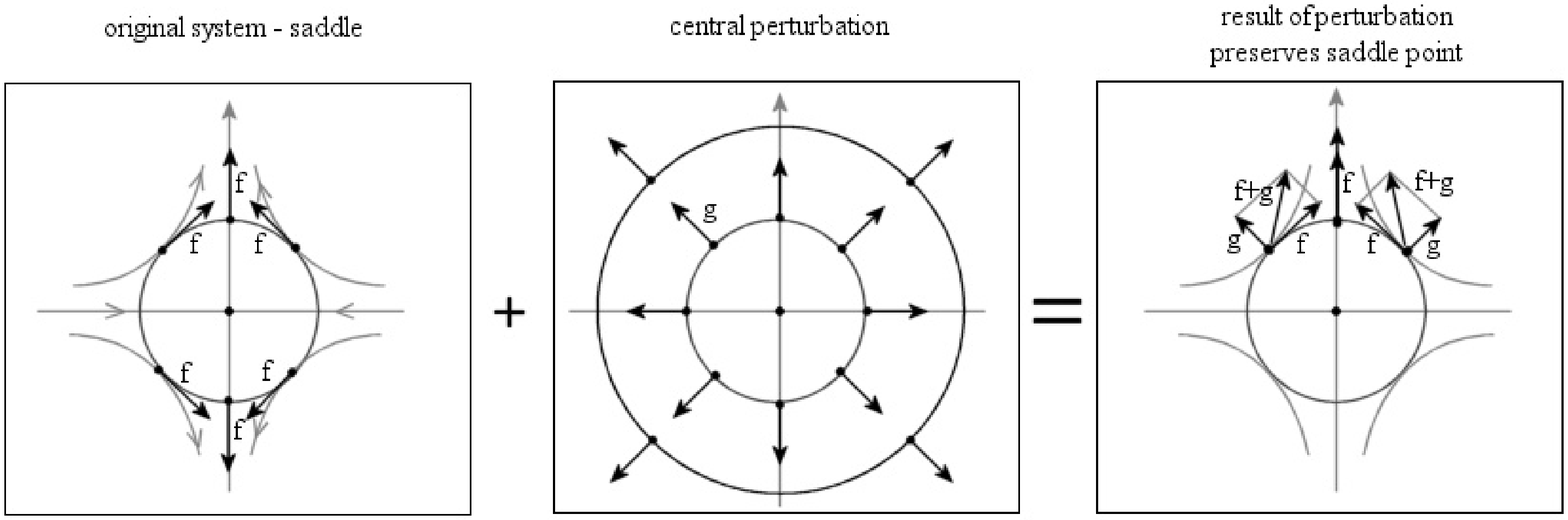}

\caption{Illustration of the structural instability of center and (b)
structural stability of saddle point. A perturbation of the original vector
field $\boldsymbol{f}$ means the addition of a small vector
$\boldsymbol{g}$. The effect of perturbation modifies it at every point in the
state space (see (a)). The effect of adding a global perturbation to the original
system disturbs a center point (see (b)) and preserves a saddle point as
illustrated with figure (c).}
\label{fig:1}
\end{figure}

Fig.~\ref{fig:1} illustrates the property of structural stability of a single
spiral attractor (focus) and saddle point, and the structural instability of 
center. The addition of a delta perturbation pointing outward (no matter how 
week) results in a point repellor. We call such a system structurally unstable
because phase portrait of the center and focus are not topologically equivalent
(note that all phase curves around the center are closed in contrast to the
focus). Hence one can claim that the pendulum system (without friction) is
structurally unstable.

The idea of structural stability attempts to define the notion of stability of
differential deterministic models of physical processes.

For planar dynamical systems (as is the case for the models under 
consideration) Peixoto's theorem \cite{Peixoto:1962ss} states that structurally 
stable dynamical systems form open and dense subsets in the space of all 
dynamical systems defined on the compact manifold. This theorem is a basic 
characterization of the structurally stable dynamical systems on the plane 
which offers the possibility of an exact definition of generic (typical) and
non-generic (exceptional) cases (properties) employing the notion of structural
stability. Unfortunately, there are no counterparts of this theorem in more
dimensional cases when structurally unstable systems can also form open and
dense subsets. For our aims it is important that Peixoto's theorem can
characterize generic cosmological models in terms of the potential function.

While there is no counterpart of Peixoto's Theorem in higher dimensions, it is
easy to test whether a planar polynomial system has a structurally stable
global phase portrait. In particular, a vector field on the Poincar{\'e} sphere
will be structurally unstable if there is a non-hyperbolic critical point at
infinity or if there is a trajectory connecting saddles on the equator of the
Poincar{\'e} sphere $S^{2}$. In opposite case if additionally the number of
critical points and limit cycles is finite, $\boldsymbol{f}$ is structurally
stable on $S^{2}$ (see \cite[p. 322]{Perko:1991de}). Following Peixoto's 
theorem the structural stability is a generic property of the $C^{1}$ vector 
fields on a compact two-dimensional differentiable manifold $\mathcal{M}$.

Let us introduce the following definition
\begin{definition}
If the set of all vector fields $\boldsymbol{f} \in C^{r}(\mathcal{M})$ 
($r \ge 1$) having a certain property contains an open dense subset of
$C^{r}(\mathcal{M})$, then the property is called generic.
\end{definition}

From the physical point of view it is interesting to known whether a certain
subset $V$ of $C^{r}(\mathcal{M})$ (representing the class of cosmological
accelerating models in our case) contains a dense subset because it means
that this property (acceleration) is typical in $V$ (see Fig.~\ref{fig:1}).

It is not difficult to establish some simple relation between the geometry of
the potential function and localization of the critical points and its character for
the case of dynamical systems of the Newtonian type:
\begin{enumerate}
\item The critical points of the systems under consideration $\dot{x}=y$,
$y=-\frac{\partial V}{\partial x}$ lie always on the $x$ axis, i.e. they
represent static universes $y_{0}=0$, $x=x_{0}$;
\item The point $(x_{0},0)$ is a critical point of the Newtonian system iff it
is a critical point of the potential function $V(x)$, i.e. $V(x)=E$ ($E$ is the 
total energy of the system $E=y^{2}/2 + V(x)$; $E=0$ for the case of flat 
models and $E=-k/2$ in general);
\item If $(x_{0},0)$ is a strict local maximum of $V(x)$, it is a saddle type 
critical point;
\item If $(x_{0},0)$ is a strict local minimum of the analytic function $V(x)$, 
it is a center;
\item If $(x_{0},0)$ is a horizontal inflection point of the $V(x)$, it is a 
cusp;
\item The phase portraits of the Newtonian type systems have reflectional
symmetry with respect to the $y$ axis, i.e. $x \to x$, $y \to -y$.
\end{enumerate}

\begin{figure}
\includegraphics[scale=1]{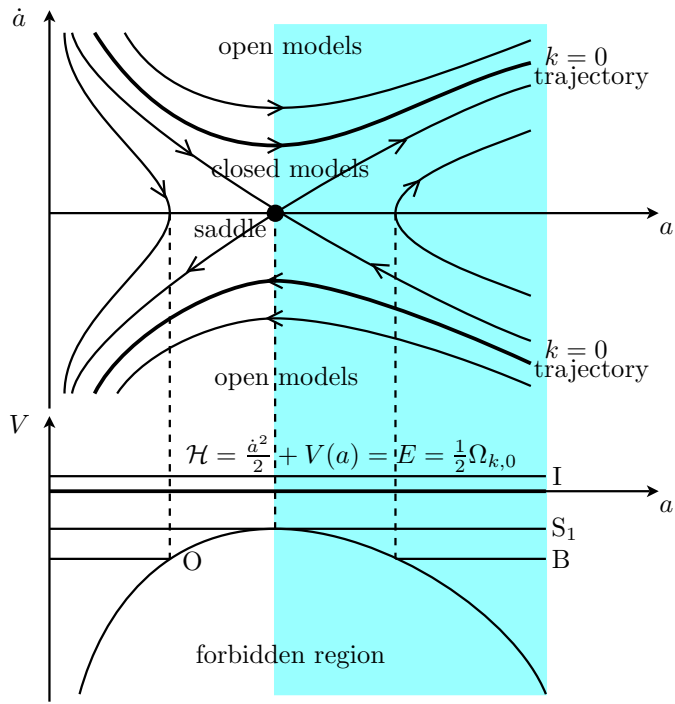}
\caption{The model of an accelerating universe given in terms of the potential 
function and its phase portrait. The domain of acceleration is represented by 
shaded area. Note that only in this case with a single 
maximum of the potential function there are two phases of the universe evolution 
where the acceleration phase follows the deceleration phase. Note also, that 
the order of appearing of acceleration epochs is important and only in this 
case we obtain a structurally stable phase portrait. Therefore, from
Peixoto's theorem we obtain the generic phase portrait for the accelerating
universe. It is equivalent to the $\Lambda$CDM model scenario.}
\label{fig:2}
\end{figure}

All these properties are simple consequences of the Hartman-Grobman theorem
which states that near the non-degenerate critical points (hyperbolic) the
original dynamical system is equivalent to its linear part. Therefore the
character of a critical point is determined by the eigenvalues of the 
linearization matrix given by a simple equation $\lambda^{2} + \det{A} = 0$, 
where 
\[
A= \left[ \begin{array}{cc}
0 & 1 \\
-\frac{\partial^{2} V}{\partial x^{2}} & 0 
\end{array} \right]_{(x_{0},0)}
\]
is the linearization matrix. Hence, in the case of a maximum we obtain a saddle 
with $\lambda_{1}$, $\lambda_{2}$ real of opposite signs, and if the potential 
function assumes a minimum at the critical point we have a center with 
$\lambda_{1,2}$ purely imaginary of mutually conjugate. Therefore, among all
distinguished cases, only if the potential function admits a local maximum at 
the critical point we have a structurally stable global phase portrait. Because 
$V \le 0$ and $\partial V/ \partial a = \frac{1}{6} (\rho + 3 p)(a)$ the 
Universe is decelerating if the strong energy condition is satisfied and 
accelerating if the strong energy condition is violated. Hence, among all 
simple scenarios, the one in which deceleration is followed by acceleration is 
the only structurally stable one (see Fig.~\ref{fig:2}).

\begin{figure}
\includegraphics[scale=1]{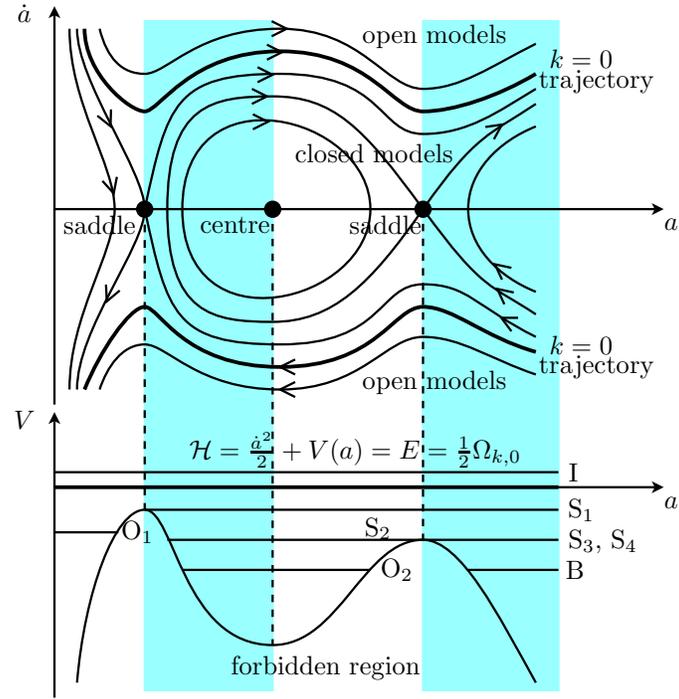}
\caption{The model of an accelerating universe given in terms of the potential 
function and its phase portrait. The domain of acceleration is represented by 
shaded area. In this case the existence of two maxima 
induces the appearance of minima of the potential function whose presence gives 
rise to structural instability. Therefore, from Peixoto's theorem we obtain 
the non-generic phase portrait for the universe accelerating in two domains.}
\label{fig:3}
\end{figure}

\begin{figure}
\includegraphics[scale=1]{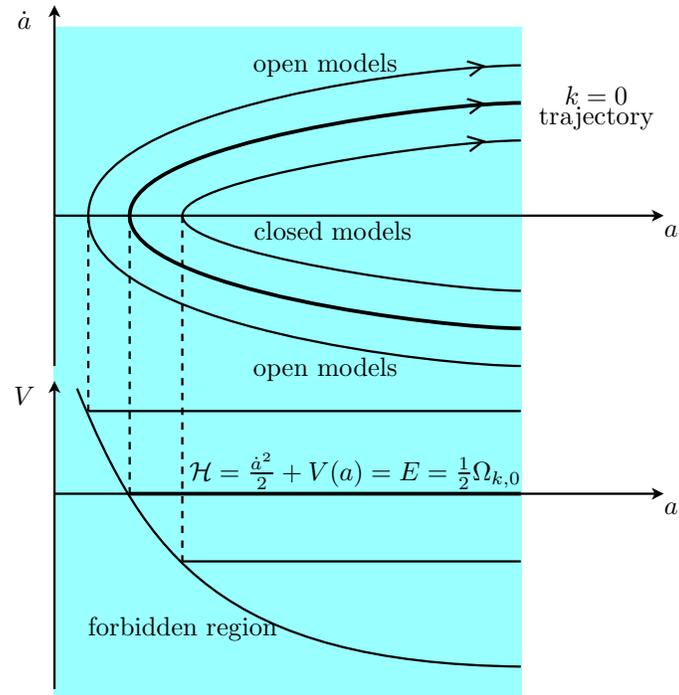}
\caption{The model of an accelerating universe given in terms of the potential 
function and its phase portrait. The domain of acceleration is represented by 
shaded area. In this case the universe is accelerating for all trajectories. 
In the phase portrait there is no critical point in the finite domain. The 
potential function is decreasing function of the scale factor. While this 
system is structurally stable there is no matter dominating phase in the model. 
And from physical view of point is not interesting.}
\label{fig:4}
\end{figure}

\begin{figure}
\includegraphics[scale=1]{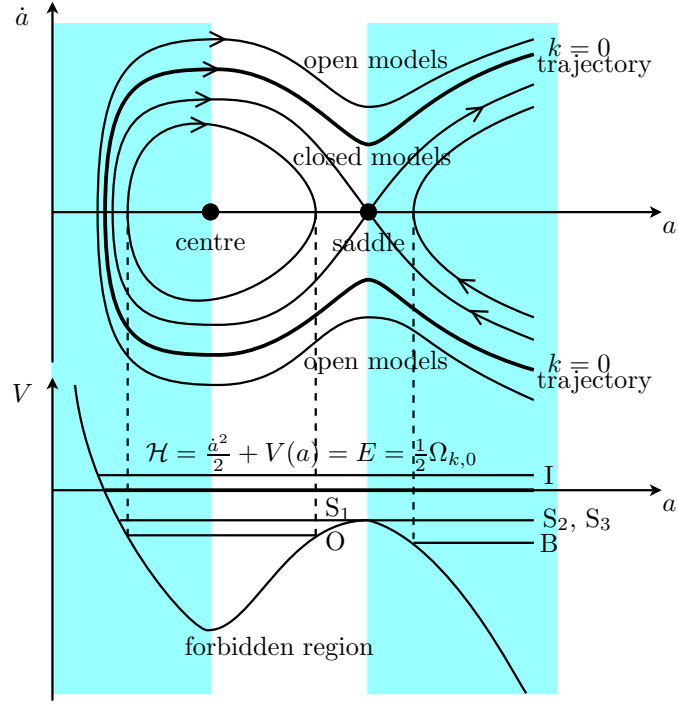}
\caption{The model of an accelerating universe given in terms of the potential 
function and its phase portrait. The domain of acceleration is represented by 
shaded area. The phase portrait of extended bouncing models with the 
cosmological constant. While the early evolution is dominated by matter terms, 
for late times the cosmological term is dominating. There are three 
characteristic types of evolution: I -- inflectional, O -- oscillating, B -- 
bouncing. The model is structurally unstable because of the presence of a 
non-hyperbolic critical point (center) on the phase portrait.}
\label{fig:5}
\end{figure}

\begin{figure}
\includegraphics[scale=1]{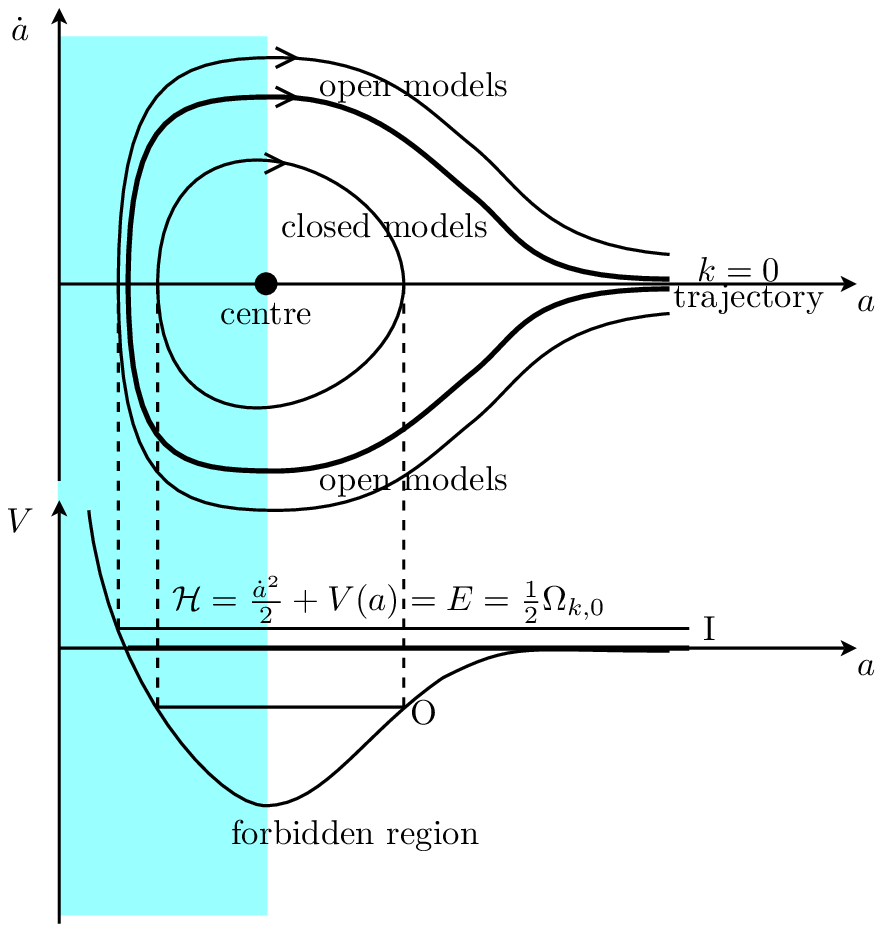}
\caption{The model of an accelerating universe given in terms of the potential 
function and its phase portrait. The domain of acceleration is represented by 
shaded area. The phase portrait as well as the diagram of the corresponding 
potential for bouncing models $H^{2}/H_{0}^{2} = \Omega_{\text{m},0}x^{-m} 
- \Omega_{n,0}x^{-n}$, $n>m$, $m,n = \text{const}$. All trajectories represent 
cosmological models in which the phase of expansion is proceeded by the phase of 
contraction. They are closed for oscillating (O) models appearing for positive 
curvature, or open, representing bouncing flat and open cosmologies with a 
single bounce phase. On the phase portrait we find a single critical point of 
the center type. Because the presence of a critical point of non-hyperbolic 
type, the system is structurally unstable. The motion of the system is 
determined in the domain of the configuration space $V(a)<0$, i.e. for 
$a \ge a_{\text{min}}$ in the case presented on the figure. Note that for 
oscillating models there is an infinite number of transitions from acceleration 
to deceleration epoch. For flat and open cosmologies there is only one such 
transition.}
\label{fig:6}
\end{figure}

Let us consider two types of scenarios of cosmological models with matter 
dominated and dark energy dominated phases
\begin{enumerate}
\item the $\Lambda$CDM scenario, where the early stage of evolution is 
dominated by both baryonic and dark matter, and late stages are described by 
the cosmological constant effects.
\item the bounce instead initial singularity squeezed into a cosmological 
scenario; one can distinguish cosmological models early bouncing phase of 
evolution (caused by the quantum bounce) \cite{Singh:2006im} from the classical 
bouncing models at which the expansion phase follows the contraction phase. In 
this paper by bouncing models we understand the models in the former sense 
(the modern one).
\end{enumerate}

For the first class of models we obtain a global phase portrait equivalent to
that which is demonstrated in Fig.~\ref{fig:2}. The eigenvalues of this system
are real of opposite signs and this point is a saddle. It describes a stationary
but unstable universe which is quite similar to the static Einstein Universe
filled with both dust matter and with cosmological constant. There are, in
principle, three representative scenarios of evolution. The trajectories moving
in the region $B$ confined by the separatrices correspond to the closed
universes contracting from the unstable de Sitter node towards the stable de
Sitter node. They are sometimes called bouncing models (it is not the sense of
bouncing models in the sense used by us in this paper -- see Tables \ref{tab:1} 
and \ref{tab:2} for comparison). All trajectories on the phase plane are 
divided by a parabolic curve (representing the trajectory of the flat model) 
into two disjoint classes of closed and open models. The trajectory situated 
in the region $I$ confined by the upper branch of the trajectory $k=0$ and by 
the separatrices (coming out and approaching the saddle) of the saddle point, 
correspond to the closed universes expanding from the initial singularity 
($x=0$, $\dot{x}=\infty$) to the stable deS$_{+}$ attractor. Quite similarly, 
the trajectories located in the symmetric region $x \to x$ and $y \to -y$ 
correspond to the closed contraction universes ($y < 0$) from the unstable de 
Sitter node towards the singularity. The region $O$, whose boundaries are the 
separatrices coming out from the saddle and going to the saddle in the region 
$x \ge 0$, is covered by the trajectories of closed models which begins its 
expansion from the initial singularity, reach a maximal size and then 
recollapse to the final singularity. The trajectories situated over the upper 
branch of the trajectory of the flat model $k=0$ describe the open models 
expanding towards stable de Sitter universe from the initial singularity. They 
are called oscillating cosmological models.

The second class of models, described the cosmological models with a squeezing 
bounce phase in the cosmological scenario, is present in Fig.~\ref{fig:3}.
Its phase portrait contains a center -- non-hyperbolic critical point whose 
presence makes the system structurally unstable. Both phase portraits in 
Fig.~\ref{fig:2} and \ref{fig:3} are topologically non-equivalent in the 
sense of existence of homeomorphism. In Fig.~\ref{fig:4} we present a phase 
portraits of cosmological models with acceleration in all time; however there 
is no matter dominating phase during its evolution. In Fig.~\ref{fig:5} it is 
presented a sub-case which starts with accelerating phase. This model is 
similar to the models considered in loop quantum cosmology which started with 
acceleration at very beginning.  

In the simplest case we obtain a phase portrait equivalent to that presented on
Fig.~\ref{fig:6}. Notice that all trajectories are bouncing and trajectories
around the center are oscillating. The accelerating region is situated
left of the critical point. In Fig.~\ref{fig:5} the qualitative behaviour
of the universes whose late time evolution is dominated by the cosmological
constant term is presented. In this case we have two disjoint acceleration areas
corresponding to that domain of configuration space at which the potential is
a decreasing function of the scale factor.

One can imagine different evolutional scenarios in terms of the potential 
function (see Fig.~\ref{fig:2}-\ref{fig:6}). Because of the existence of a 
bouncing phase which always gives rise to the presence of a non-hyperbolic 
critical point on the phase portrait one can conclude
\begin{itemize}
\item the bounce is not a generic property of the evolutional scenario,
\item structural stability prefers the simplest evolutional scenario in which
the deceleration epoch is followed by the acceleration phase.
\end{itemize}
The dynamical systems with the property of such switching rate of expansion, 
following the single-well potential are generic in the class of all dynamical 
systems on the plane.

The presented approach to describe dynamics can be extended to the case of
cosmological models with scalar field. They play an important role in the 
quintessence conception. To illustrate this let us consider a homogeneous, 
minimally coupled scalar field on the FRW background. The dynamical effects of 
this scalar field are equivalent to the effects of a perfect fluid with energy 
density and pressure given in the form
\[
\begin{array}{l}
\rho_{\phi} = \frac{1}{2} \varepsilon \dot{\phi}^{2} + U(\phi),\\
p_{\phi} = \frac{1}{2} \varepsilon \dot{\phi}^{2} - U(\phi),
\end{array}
\]
where $\varepsilon = +1$ for standard scalar field and $\varepsilon = -1$ for
phantom scalar field, $U(\phi)$ is the potential of a scalar field.

A construction analogous to the presented above is to use the expression for
the effective energy density. Let us consider for example a universe filled
with
perfect fluid with pressure $p=w \rho$, $w= \text{const}$ and a minimally
coupled scalar field. Then, we can adopt the standard formula
$V=-\frac{\rho_{\text{eff}}a^{2}}{6}$ (if we use conformal time $\ud \eta =
\ud t/a$ then $V=-\frac{\rho_{\text{eff}}a^{4}}{6}$) and we obtain
\[
V = -\frac{1}{6} \rho_{w,0}a^{-1-3w} -\frac{1}{6} \rho_{\phi}a^{2}.
\]
After substituting $\rho_{\phi}$ into the above formula and the shifted kinetic
term $-\frac{\varepsilon}{12}\phi^{2}$ into the kinetic energy of the system
(remember that the division into the kinetic and potential parts has a purely
conventional character), we obtain a $2$-dimensional Hamiltonian system in the 
form 
\begin{equation}
\label{eq:d1}
\mathcal{H} = \frac{1}{2}\left(\frac{\ud a}{\ud t}\right)^{2} 
- \frac{1}{2} \varepsilon a^{2} \left(\frac{\ud \phi}{\ud t}\right)^{2} 
+ \psi(a,\phi) \equiv -\frac{k}{2},
\end{equation}
where in above formula we use units in which $8 \pi G=c=1$ (some authors
put $4\pi G/3 = c=1$, then $V\equiv - \rho_{\text{eff}}a^{2}$), $\phi$
is a rescaled function $\phi \to \phi/\sqrt{6}$ and
\[
\psi(a,\phi) = a^{2} U(\phi) + \frac{1}{6}\rho_{w,0} a^{1-3w}.
\]
In the special case of the radiation filled universe ($w=1/3$) we obtain
Hamiltonian system defined on the zero energy level
\[
E \equiv -\frac{k}{2} + \frac{1}{6} \rho_{r,0} = \text{const}.
\]
Therefore, in the case of potential $U(\phi)=\frac{1}{2}m^{2}\phi^{2} + \lambda
\phi^{4}$ we obtain the $2$-dimensional potential function $\psi(a,\phi)$ which 
identify quintessence model. The same approach can be adopted to the case of 
conformally coupled scalar field a well as for complex scalar fields.

It is interesting to investigate what a class of quintessence models which 
reduces the dynamics of the universe with minimally coupled scalar field 
$\Phi$ (evolving in the potential $U(\Phi)$) to the dynamical system of 
a Newtonian type with the potential parameterized by the scale factor 
$V(a)$. The quintessence model is determined completely once the potential 
function $U$ of the scalar field is given. Thus energy density and pressure 
is given by
\[
\rho_{\text{eff}} = -6 \frac{V(a)}{a^2}.
\]

On the other hand the dynamical system of a Newtonian type is specified 
completely once the potential $V(a)$ is fixed. Therefore if we assume 
that $\Phi = \Phi(a(t))$ then 
\begin{align*}
\rho_{\Phi} &= \frac{1}{2} \Phi'(a)^2 a^2 H^2(a) + U(\Phi(a)) \\
p_{\Phi} &= \frac{1}{2} \Phi'(a)^2 a^2 H^2(a) - U(\Phi(a)) 
\end{align*}
and
\begin{equation*}
w_{X}(a) = \frac{\Phi'(a)^2 a^2 H^2(a) - 2U(\Phi(a))}{\Phi'(a)^2 a^2 H^2(a) 
+ 2U(\Phi(a))}
\end{equation*}
where $a^2 H^2(a) = - 2V(a)$. 

Hence the particle-like approach can be extended naturally on the class of 
phenomenological quintessence models with the parameterized equation of state 
by redshift (or the scale factor equivalently) 
\cite{Linder:2002et,Seljak:2004xh}. In this approach it is assumed the equation 
of state $p_{X} = w_{X}(a) \rho_{X}$ and $w_{X} = w_{X}(a)$ such that 
\[
\rho_{X} = \rho_{X,0} a^{-3(1+\bar{w}_{X}(a))} = 
\rho_{X,0} a^{-3} e^{\ln a^{-3\bar{w}_{X}(a)}} = 
\rho_{X,0} a^{-3} e^{-3\int_{1}^{a} w_{X}a^{-1} da}
\]
where $\bar{w}_{X}(a)$ is the mean of the equation of state in the 
logarithmic scale, i.e.
\[
\bar{w}_{X}(a) = \frac{\int w_{X}(a) d(\ln a)}{\int d\ln a}
\]
The main motivation of the above assumption is the explanation of cosmic 
coincidence problem: why do all the contributions from the vacuum energy 
density (the cosmological constant) are comparable with energy density of 
matter? To remove the fine-tuning problem it was proposed a simple power law 
relation $\bar{w}_{X}(a) = w_{0} a^{\alpha}$, $w_{0}, \alpha = \text{const}$ 
(scalling fluid) \cite{Rahvar:2006tm}. If we consider a scalar field 
parameterized by the scale factor, the class of possible quintessence paths 
is restricted from definition. Note that cosmography measures only average 
properties of matter expressed in kinematics of $H(z)$ and from (1) we 
obtain \cite{Guo:2005at}
\begin{align}
\label{eq:d2}
\left( \frac{d\Phi}{dt}\right)^2 &= \rho_{X}(1+w_{X}) \\
U(\Phi) &= \frac{1}{2} \rho_{X}(1-w_{X})
\end{align}
where it is assumed that both the scalar field $\Phi$ and its potential 
$V(\Phi)$ depends on time through the scale factor, i.e. $\Phi(t) = 
\Phi(a(t))$, $U(\Phi(t)) = U(\Phi(a))$. Then equation (\ref{eq:d2}) can 
be rewritten to the new form
\begin{equation}
\label{eq:d3}
\left( \frac{d\Phi}{da}\right)^2 = \frac{1}{a^2 H^2} \rho_{X}(a)[1+w_{X}(a)]
\end{equation}
or in terms of redshift $z$ 
\begin{equation}
\label{eq:d4}
\left( \frac{d\Phi}{dz}\right)^2 = - \frac{\rho_{X}(z)[1+w_{X}]}{H^2(z)[1+z]^2} 
\quad \Longrightarrow \quad \Phi = \Phi(z)
\end{equation}
where we use the standard relation $1+z=a^{-1}$ and $a_0 = 1$ is the value 
of the scale factor at present epoch. The potential $V$ in the dynamical 
system under consideration and the potential of the scalar field can be 
written as 
\begin{equation}
V(a(z)) = - \frac{a^2}{6} \left[ \rho_{\text{m},0} a^{-3} + \rho_{X,0} 
a^{-3(1+\bar{w}_{X})}\right]
\end{equation}
or
\begin{equation}
V(a(z)) = - \frac{1}{6} \left[ \rho_{\text{m},0} a^{-3} + \rho_{X,0} 
a^{-3} e^{-3\int_{1}^{a} w_{X}(a) d(\ln a)}\right] a^2
\end{equation}
and
\begin{equation}
U(a) = \rho_{X,0} a^{-3} e^{-3\int_{1}^{a} w_{X}(a) d(\ln a)} (1-w_{X})
\end{equation}
Hence
\begin{equation}
V(a) = - \frac{1}{6} \left[ \rho_{\text{m},0} a^{-3} + (1-w_{X})^{-1} U(a) 
\right] a^2.
\end{equation}

Let us return to the general Hamiltonian (\ref{eq:d1}) then after substitution 
(\ref{eq:d3}) to the potential $V=- \rho_{\text{eff}}a^2/6$, $\rho_{\text{eff}} 
= \rho_{\Phi} + \rho_{\text{m}}$
\begin{align}
V &= - \frac{1}{6} a^{2} \left[ \frac{\rho_{X}(a)(1+w_{X}(a))}{2} + U(a) \right] 
- \frac{1}{6} \rho_{\text{m},0} a^{-1} 
= - \frac{1}{6} a^{2} \left[ \frac{\rho_{X}(a)(1+w_{X}(a))}{2} 
+ \frac{1}{2}\rho_{X}(1-w_{X})  \right] - \frac{1}{6} \rho_{\text{m},0} a^{-1} 
\nonumber \\
&= -\frac{1}{6} \rho_{X}(a) a^{2} - \frac{1}{6} \rho_{\text{m},0} a^{-1} 
\equiv - \frac{1}{6} \rho_{\text{eff}} a^{2}
\end{align}

This means that after parameterization by the scale factor the scalar field 
quintessence model has the potential in the form just prescribed in the 
particle like description of quintessential models in which the phase space 
is two-dimensional instead of 4-dimensional ($\Phi, \dot{\Phi}, a, \dot{a}$) 
\cite{Szydlowski:2003fg,Szydlowski:2003cf}.
Let us note that in this case the potential reconstruction performed by 
Rahvar and Mohaved is equivalent to the reconstruction of the 
corresponding potential function of the dynamical system $V(a)$. All dynamics 
is now squeezed to the plane $(a, \dot{a})$.

\section{The value of transition redshift and the Hubble parameter at the 
transition epoch}
\label{sec:3}

Due to the existence of a simple relation between the luminosity distance
$d_{L}(z)$ as a function of redshift $z$ and the Hubble function
\begin{equation}
H(z) = \left[ \frac{\ud}{\ud z}\left(\frac{d_{L}(z)}{1+z}\right) \right]^{-1},
\label{eq:10}
\end{equation}
it is possible to reconstruct the potential function $V(a)$. Note that even if
the luminosity distance i obtained accurately the potential cannot be
determined uniquely as it depends on $\Omega_{k,0}$ -- an additional curvature
parameter.

If we assume that the model is flat, then the trajectories of the corresponding
Hamiltonian system lies on the zero energy level and
\begin{equation}
V(z) = -\frac{1}{2} \left\{(1+z) \frac{\ud}{\ud z} 
\left( \frac{d_{L}(z)}{1+z}\right) \right\}^{-2}.
\label{eq:11}
\end{equation}

\begin{figure}
\includegraphics[scale=0.5]{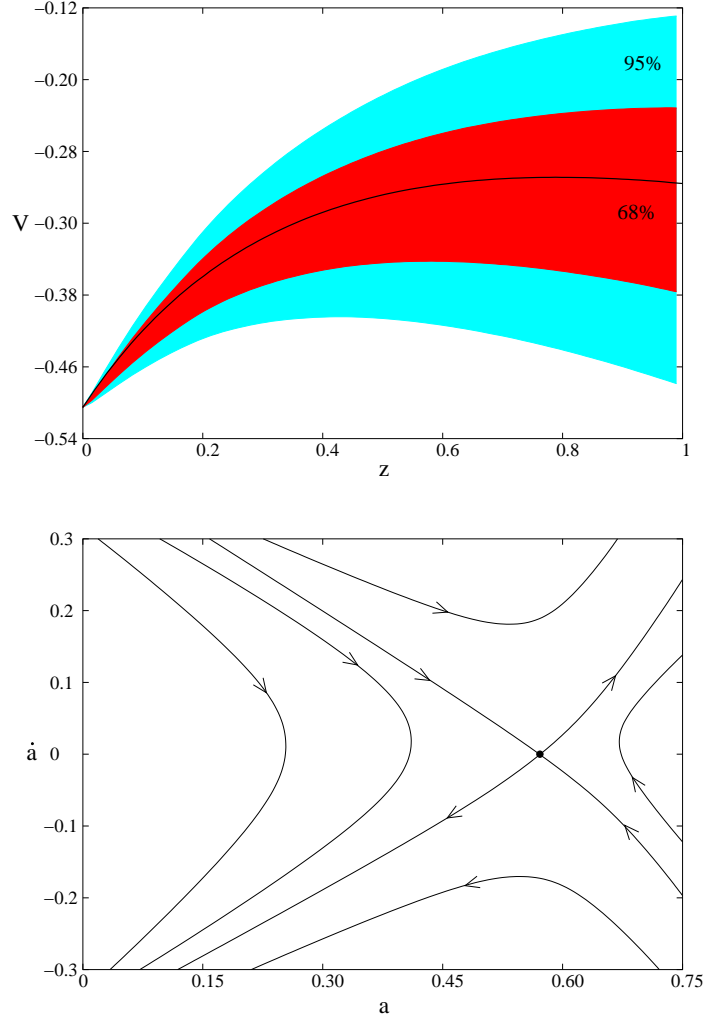}
\caption{The potential function of the system reconstructed from SNIa data 
(the Riess et al. sample).}
\label{fig:7}
\end{figure}

The result of this reconstruction from the Riess sample is shown in 
Fig.~\ref{fig:7}. From this figure we obtain the shape of the potential function 
like the one of the $\Lambda$CDM model. Hence the corresponding phase portrait 
contains a single maximum. The value of redshift at this point we call the 
transition redshift and denote $z_{T}$. It is possible to obtain not only the 
qualitative shape of the potential function (an inverted single-well potential) 
but also some quantitative attributes of the model. Especially it is well to 
know when the switch from the deceleration to the acceleration epoch occurs in 
different cosmological scenarios. The result of fitting of the Newtonian type 
system with the simple potential function given in the simple polynomial form is 
presented in Table~\ref{tab:2}. From this estimation we obtain $z_{T}\simeq 0.5$ 
which can be interpreted as the value of redshift $z_{T}$ at which the switch 
between deceleration and acceleration epochs takes place.

In Fig.~\ref{fig:7} one can see that the universe is accelerating in the 
redshift interval in which the potential function decreases in respect to $z$. 
On the other hand the acceleration takes place in the region where the strong 
energy condition is violated. The idea of testing of energy condition through 
the measurement of distant SNIa was done by Santos et al. \cite{Santos:2006ja}. 

Due to the particle like description dynamics of accelerating models one can 
also estimate the value of the Hubble function at the moment of the transition 
redshift, namely
\begin{equation}
H^{2}(z_{T}) = -2 (1+z_{T})^{2} H_{0}^{2} V(z_{T}).
\label{eq:12}
\end{equation}

From equation~(\ref{eq:12}) we obtain that
\begin{equation}
H(z_{T}) \simeq 0.9 H_{0}.
\label{eq:13}
\end{equation}
There arises a basic problem in connection with the above coincidence: Why the 
value of $H(z_{T})$ and the present value of the Hubble function $H_{0}$ are of 
the same order of magnitude? The nature of this problem is similar to the 
cosmic coincidence conundrum \cite{Gorini:2004by}.

From the definition of the potential function we can obtain a simple
interpretation of acceleration of the Universe. For example, the deceleration
parameter at the present epoch is the slope of the potential ($q_{0}=(\partial
V/\partial a)_{a=1}$), jerk $j=\dddot{a} a^{2}/\dot{a}^{3}$ is related to the
convexity of $V(x)\colon j_{0}=-(\partial V/\partial a^{2})_{a=1}$. Our results
\cite{Szydlowski:2005,Szydlowski:2004np} also show the advantage of using 
$V(x)$ instead of the coefficient of the equation of state $w_{X}(z)$ to probe 
the variation of dark energy.

It have been recently shown the advantages of using $\rho_{X}(z)$ and $H(z)$, 
instead of the coefficient of equation of state \cite{Wang:2004ru,Wei:2006ut}. 
Note that both potential approaches are useful to differentiate between 
different dark energy propositions.

The potential function in the neighbourhood of its maximum can be approximated by 
\[
V(a) = V(a_{T}) + \left(\frac{\ud^{2} V}{\ud a^{2}}\right)_{a_{T}} 
\frac{(a-a_{T})^{2}}{2}.
\]
Hence we can calculate the energy density near the transition epoch $a=a_{T}$
\[
\rho(a) = -\frac{6}{a^{2}}(a_{T}^{2}+V(a)) 
\left(\frac{\ud^{2} V}{\ud a^{2}}\right)_{a_{T}} 
- 3 \left(\frac{\ud^{2} V}{\ud a^{2}}\right)_{a_{T}} 
+ 6\frac{a_{T}}{a}\left(\frac{\ud^{2} V}{\ud a^{2}}\right)_{a_{T}},
\]
where $\left(\frac{\ud^{2} V}{\ud
a^{2}}\right)_{a=a_{T}}=\frac{1}{6}a_{T}\frac{\ud}{\ud a}|_{a_{T}}(\rho+3p)$.
The first and second terms are positive and correspond to $1$-dimensional 
topological defects and positive cosmological constant terms. The third term 
is negative and scales like $2$-dimensional topological defects.

It is useful to define space time metric of the non flat universe in the new 
variables
\[
\ud s^{2} = a^{2} \left\{\left[\frac{\ud (\ln{a})}{\sqrt{-2V}}\right]^{2} 
-\ud \Omega^{2}_{3}\right\}
\]
or
\[
\ud s^{2} = a^{2} \left\{ \ud \tau^{2} - \ud \Omega^{2}_{3} \right\},
\]
where $\ud \tau = \frac{\ud (\ln{a})}{\sqrt{-2V}}$. Therefore, the only 
nontrivial metric function in a FRW cosmology is the function of $V(a)$ and 
the value of curvature which is encoded in the spatial part of the line 
element. Hence one can conclude any kind of observation based on geometry 
(cosmography) will allow us to determine a single potential function (for 
comparison see \cite{Padmanabhan:2002ji}). As argued by Padmanabhan, this 
function is insufficient to describe matter content of the universe and some 
additional input is still required.

Let us consider now the general properties of dark energy dynamics in terms of
the potential. Due to particle-like description of cosmology with dark energy, 
the methods of qualitative analysis of differential equations can be naturally
adopted. The main advantage of this method is the possibility of investigating
all admissible evolutional paths for all initial conditions in the geometrical
way -- on the phase plane. The structure of the phase space is organized by
singular solutions of the system which are represented by critical points
(points of the phase plane for which right hand sides of the system vanishes) 
and phase curves connecting them. Following the Hartman-Grobman theorem the 
behaviour of the trajectories near the critical points is equivalent to the 
trajectories of linearized system at this point. Therefore for constructing of 
the picture of global dynamic called a phase portrait it is necessary to 
investigate all critical points and their type (determine the stability). The
critical points $(x_{0},0)$ correspond to $y_{0}=0$ and $(\partial V/\partial
x)_{x_{0}}=0$. They are saddle point if $V_{xx}(x_{0})<0$ and then eigenvalues
of the linearization matrix $\lambda_{1,2}=\pm \sqrt{-V_{xx}(x_{0})}$ are real
of opposite signs or centres if $V_{xx}(x_{0})>0$ and then eigenvalues are
purely imaginary $\lambda_{1,2}=\pm i \sqrt{V_{xx}(x_{0})}$. The exceptional
case of $V_{xx}(x_{0})=0$ is degenerate. Because only static critical point
are admissible due to the constraint condition (Friedmann first integral) we 
obtain that $V(x_{0})=\frac{1}{2} \Omega_{k,0}$. Therefore, because 
$V(x) \le 0$, the critical points are admissible for $\Omega_{k,0} \le 0$, 
i.e. for the closed model only.

The system linearized around the critical point $x_{0}$ has the form
\begin{align}
(x-x_{0})\dot{} &= (y-0),\\
(y-y_{0})\dot{} &= \left(-\frac{\partial^{2} V}{\partial x^{2}}\right)_{x=x_{0}} (x-x_{0}),
\end{align}
which is equivalent to a single differential equation of the second order
\[
(x-x_{0})\ddot{} = \left(-\frac{\partial^{2} V}{\partial x^{2}}\right)_{x=x_{0}} (x-x_{0}).
\]
The solution of the above linear system which approximates the behaviour of
the trajectories near the critical point is
\[
x-x_{0}=\frac{\dot{x}_{0}}{\sqrt{-(\partial^{2} V/\partial x^{2})_{x_{0}}}} 
\sinh{\sqrt{\left(-\frac{\partial^{2} V}{\partial x^{2}}\right)_{x_{0}}}(t-t_{0})},
\]
where $x_{0}=x(t_{0})$ and $\dot{x}_{0}$ can be determined from the Friedmann
first integral $\dot{x}_{0}=\sqrt{-2 V(a_{0})}x_{0}$. The special choice
$t_{0}=\pm \infty$ corresponds to the separatrices in and out-going from the
static critical point.

Let us consider now the case of a saddle point. Then the angle of slopes of 
the separatrices at this point 
\[
\tan^{2}{\frac{\alpha}{2}} = 
- \left(\frac{\partial^{2} V}{\partial x^{2}}\right)_{x_{0}},
\]
where $\alpha$ is the angle between eigenvectors at the critical point of the 
saddle type.

On the other hand from the reconstructed form of the potential function one can
determine the dynamics on the phase plane without any information about the 
value of $\Omega_{k,0}$. Hence, we can establish the angle $\alpha$. It is 
strictly related to the value of jerk because of the relation 
$j_{0}=-(\partial^{2} V/\partial x^{2})_{x_{0}}$. Finally we obtain
\[
\alpha = 2 \arctan{j_{0}}.
\]

The universe is accelerating in such a domain of the configuration space in
which $V(x)$ is a decreasing function of its argument. One can calculate
the average value of time which trajectories spend during the loitering epoch
($\dot{a}$ close to zero)
\[
\Delta t = \frac{1}{2 x_{0}} \int^{x_{0}+\Delta x}_{x_{0}
-\Delta x} \frac{x \ud x}{\sqrt{-2V(X)}} 
= \frac{1}{2 x_{0}} \int^{x_{0}+\Delta x}_{x_{0}-\Delta x} \frac{\ud x}{H(x)},
\]
where $x_{0}$ is the value of the scale factor at the transition epoch 
expressed in the units of its present value.

Hence the time which the model spends in the loitering epoch depends on the 
transition epoch ($z_{\text{tr}}$) and the preassumed value of $\Delta x$ 
-- which measures deviation from this stage. For the FRW model with the 
$\Lambda$ term one can find exact forms of function $\Delta t(x_{0},\Delta x)$
in terms of the Jacobi elliptic functions.

\section{The generic and non-generic global evolutional paths in the Multiverse
of accelerating models}
\label{sec:4}

The analysis of full dynamical behaviour of trajectories requires the study of 
the behaviour of trajectories at infinity. It can be performed by means of the
Poincar{\'e} sphere construction. In this approach we project the trajectories
from center of the unit sphere $S^{2}=\{(X,Y,Z)\in \mathbf{R}^{3} \colon X^{2} 
+ Y^{2} + Z^{2}=1 \}$ onto the $(x,y)$ plane tangent to $S^{2}$ at either the
north or south pole (see \cite[p. 265]{Perko:1991de}). Due to this central
projection (introduced by Poincar{\'e}) the critical points at infinity are
spread out along the equator. Therefore if we project the upper hemisphere
$S^{2}$ onto the $(x,y)$ plane of dynamical system of the Newtonian type, then
\begin{equation}
x=\frac{X}{Z} \quad , \quad y=\frac{Y}{Z},
\label{eq:14}
\end{equation}
or
\begin{equation}
X=\frac{x}{\sqrt{1+x^{2}+\left(\frac{\partial V}{\partial x}\right)^{2}}}, 
\quad Y=\frac{y}{\sqrt{1+x^{2}+\left(\frac{\partial V}{\partial x}\right)^{2}}}, 
\quad Z=\frac{1}{\sqrt{1+x^{2}+\left(\frac{\partial V}{\partial x}\right)^{2}}}.
\label{eq:15}
\end{equation}
There is a simple way to introduce the metric in the space of all dynamical 
systems on the compactified plane.

If $\boldsymbol{f} \in C^{1}(\mathcal{M})$ where $\mathcal{M}$ is an open
subset of $\mathbf{R}^{n}$, then the $C^{1}$ norm of $\boldsymbol{f}$ can be 
introduced in a standard way
\begin{equation}
\|\boldsymbol{f}\|_{1} = \sup_{x \in E} |\boldsymbol{f}(x)| 
+ \sup_{x \in E} \| D\boldsymbol{f}(x) \|,
\label{eq:16}
\end{equation}
where $| \dots |$ and $\| \dots \|$ denotes the Euclidean norm in
$\mathbf{R}^{n}$ and the usual norm of the Jacobi matrix $D\boldsymbol{f}(x)$,
respectively.

It is well known that the set of vectors field bounded in the $C^{1}$ norm forms
a Banach space (see \cite[p. 312]{Perko:1991de}).

It is natural to use the defined norm to measure the distance between any two
dynamical systems of the multiverse. If we consider some compact subset
$\mathcal{K}$ of $\mathcal{M}$ then the $C^{1}$ norm of vector field
$\boldsymbol{f}$ on $\mathcal{K}$ can be defined as
\begin{equation}
\|\boldsymbol{f}\|_{1} = \max_{x \in \mathcal{K}} |\boldsymbol{f}(x)| 
+ \max_{x \in \mathcal{K}} \| D\boldsymbol{f}(x) \| < \infty.
\label{eq:17}
\end{equation}

Let $E=\mathbf{R}^{n}$ then the $\varepsilon$-perturbation of $\boldsymbol{f}$
is the function $\boldsymbol{g} \in C^{1}(\mathcal{M})$ form which $\|
\boldsymbol{f}-\boldsymbol{g} \| < \varepsilon$.

The introduced language is suitable to reformulate the idea of structural stability
given by Andronov and Pontryagin. The intuition is that $\boldsymbol{f}$
should be structurally stable vector field if for any vector field
$\boldsymbol{g}$ near $\boldsymbol{f}$, the vector fields $\boldsymbol{f}$ and
$\boldsymbol{g}$ are topologically equivalent. A vector field $\boldsymbol{f}
\in C^{1}(\mathcal{M})$ is said to be structurally stable if there is an
$\varepsilon > 0$ such that for all $\boldsymbol{g} \in C^{1}(\mathcal{M})$
with $\|\boldsymbol{f}-\boldsymbol{g}\|_{1} < \varepsilon$, $\boldsymbol{f}$
and $\boldsymbol{g}$ are topologically equivalent on open subsets of
$\mathbf{R}^{n}$ called $\mathcal{M}$. Note that to show that system is not
structurally stable on $\mathbf{R}^{n}$ it is sufficient to show that
$\boldsymbol{f}$ is not structurally stable on some compact $\mathcal{K}$ with
nonempty interior.

It was originally a wide spread opinion that structural stability was a typical
attribute of any dynamical system modelling adequately a physical situation. 
The $2$-dimensional case is distinguished by the fact that Peixoto's theorem 
gives the complete characterization of the structurally stable systems on any 
compact $2$-dimensional space and asserts that they form an open and dense 
subsets in the space of all dynamical systems on the plane.

Let us apply this framework to the multiverse of accelerating models 
represented in terms of a $2$-dimensional dynamical system of the Newtonian 
type on the Poincar{\'e} sphere $S^{2}$. Let the potential function be given 
in the polynomial form. If we assume that the coefficient of the equation of 
state can be expanded around the present epoch ($a=1$ or $z=0$) in the Taylor 
series, i.e.
\begin{equation}
\begin{array}{l}
p_{X}=w_{X}(a)\rho_{X},\\
w_{X}(a)=\sum_{i=0}^{N}w_{i} (1-a)^{i},
\end{array}
\label{eq:18}
\end{equation}
then we obtain from the conservation condition relation
\begin{equation}
\rho_{X}=\rho_{X,0} a^{-3(1+\sum_{i=0}^{N}(-1)^{i}w_{i})} 
\exp{\{3\sum_{k=0}^{N}(-1)^{k}\frac{(1-a)^{k}}{k} 
\sum_{i=k}^{N}(-1)^{i-k} w_{i}\}}.
\label{eq:19}
\end{equation}
Hence if we expand $\exp{\{\dots\}}$ in formula (\ref{eq:19}) then we obtain 
$\rho_{X}$ as well as $V(a)$ in the polynomial form
\begin{equation}
\rho_{X}=\rho_{X,0} a^{-3(1+\sum_{i=0}^{N}(-1)^{i}w_{i})} 
\left\{ 1+ 3\sum_{k=0}^{N}(-1)^{k}\frac{(1-a)^{k}}{k}\sum_{i=k}^{N}(-1)^{i-k} 
w_{i} + \frac{1}{2} \left[3\sum_{k=0}^{N}(-1)^{k}\frac{(1-a)^{k}}{k}
\sum_{i=k}^{N}(-1)^{i-k} w_{i}\right]^{2}\right\}
\label{eq:20}
\end{equation}
and
\begin{equation}
V(a)= -\frac{\rho_{\text{m}} a^{2}}{6} - \frac{\rho_{X}a^{2}}{6}.
\label{eq:21}
\end{equation}

In the simplest case of $w(a)$ linearized around $a=1$ ($w = w_{0}+(1-a)w_{1}$), 
we obtain the density of dark energy
\begin{equation}
\rho_{X}=\rho_{X,0} a^{-3(1+w_{0}-w_{1})} \exp{[3w_{1}(a-1)]} 
= \rho_{X,0} a^{-3(1+w_{0}-w_{1})}\left\{1+3w_{1}(a-1) 
+ 9 w_{1}^{2}(a-1)^{2}+ \cdots \right\},
\label{eq:22}
\end{equation}
where $a^{-1}=1+z$, $x=a$, $a_{0}=1$.

If we consider some subclass of dark energy models described by the vector 
field $[y,-(\partial V/\partial x)]^{T}$ on the Poincar{\'e} sphere, then the 
right hand sides of the corresponding dynamical systems are of the polynomial 
form of degree $m$. Then $\boldsymbol{f}$ is structurally stable iff (i) the 
number of critical points and limit cycles is finite and each critical point 
is hyperbolic -- therefore a saddle point in finite domain, (ii) there are no
trajectories connecting saddle points. It is important that if the polynomial
vector field $\boldsymbol{f}$ is structurally stable on the Poincar{\'e} sphere
$S^{2}$ then the corresponding polynomial vector field $[y,-(\partial
V/\partial x)]^{T}$ is structurally stable on $\mathbf{R}^{2}$
(\cite[p. 322]{Perko:1991de}). Following Peixoto's theorem the structural
stability is a generic property of $C^{1}$ vector fields on a compact
two-dimensional differentiable manifold $\mathcal{M}$. If a vector field
$\boldsymbol{f} \in C^{1}(\mathcal{M})$ is not structurally stable it belongs
to the bifurcation set $C^{1}(\mathcal{M})$. For such systems their global
phase portrait changes as vector field passes through a point in the
bifurcation set.

Therefore, in the class of dynamical systems on the compact manifold, the
structurally stable systems are typical (generic) whereas structurally unstable
are rather exceptional. In science modelling, both types of systems are used.
While the structurally stable models describe ``stable configuration''
structurally unstable model can describe fragile physical situation which
require fine tuning \cite{Tavakol:1988}.

\begin{figure}
\includegraphics[scale=1]{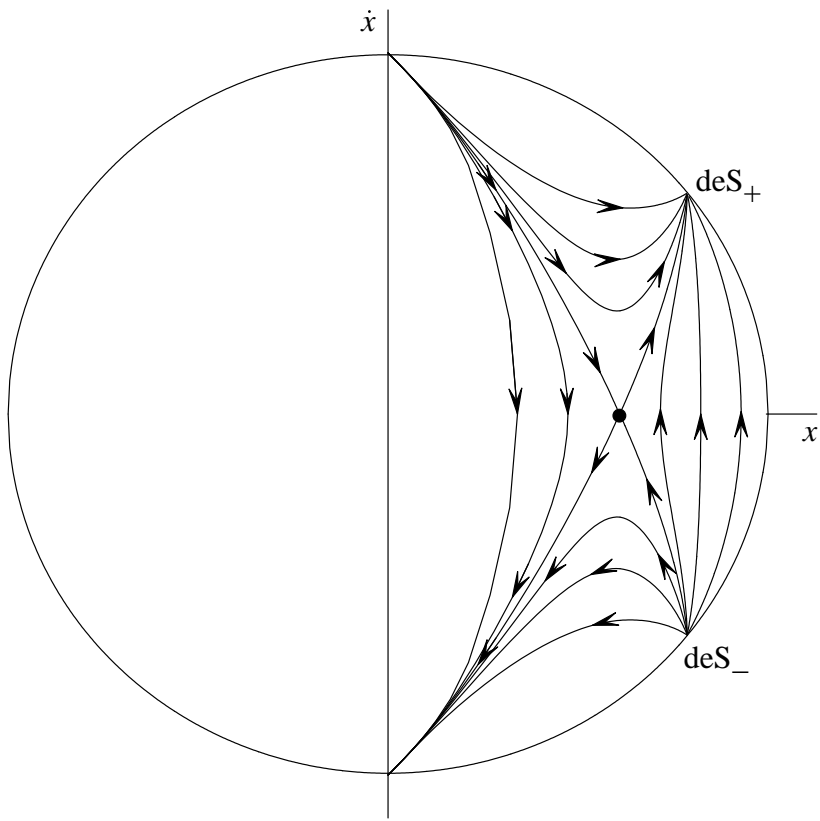}
\caption{The phase portrait of the $\Lambda$CDM model on the projective phase 
plane $\mathbf{R}P^{2}$. Using the Poincar{\'e} sphere construction we 
represent the dynamics of $\Lambda$CDM models on the compactified plane with 
the circle at infinity. Note that critical points (hyperbolic) at infinity are 
structurally stable.}
\label{fig:8}
\end{figure}

\begin{figure}
\includegraphics[scale=1]{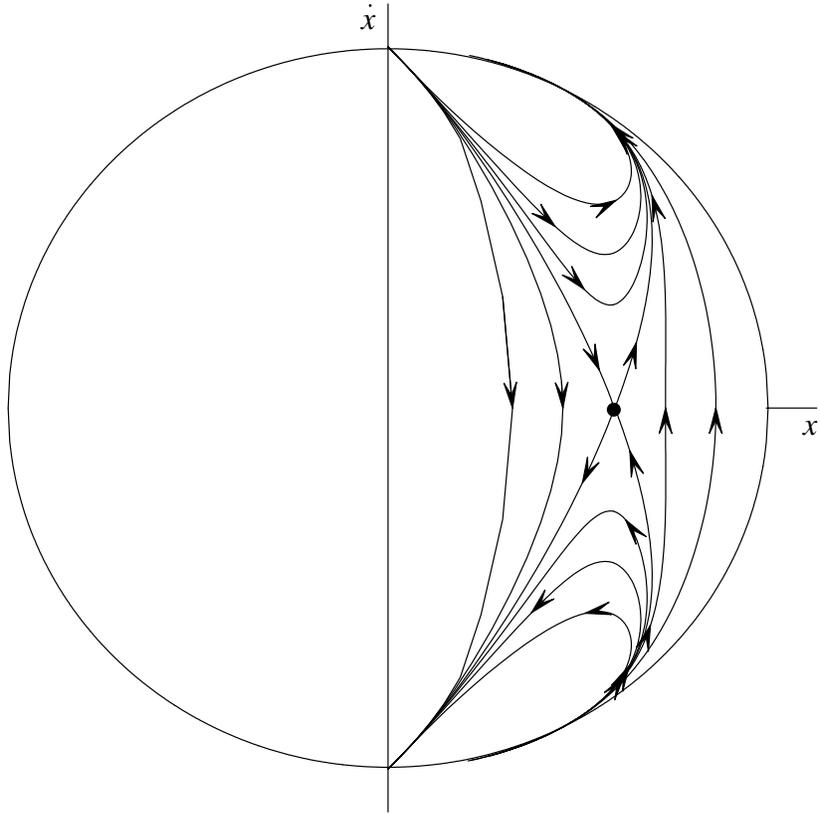}
\caption{The phase portrait for phantom 
($w_{X}<-1$) cosmology. The critical
point at the circle $x=0$, $\dot{x}=\infty$ is degenerate but after
redefinition of the positional variable $x \mapsto \bar{x}=x^{3/2}$ ($x \mapsto
\bar{x}=x^{-(1+3 w_{X})/2}$ in general) and then reparametrization of time 
following the rule $\tau \mapsto \eta \colon \frac{3}{2}\bar{x}^{1/3} \ud \tau 
= \ud \eta$, we obtain a ``regularized'' system for phantom cosmology which is 
topologically equivalent to that in Fig.~\ref{fig:8}. It is important for our 
aims that this physically equivalent dynamical system is typical in the 
multiverse.}
\label{fig:9}
\end{figure}

\begin{figure}
a)\includegraphics[scale=0.75]{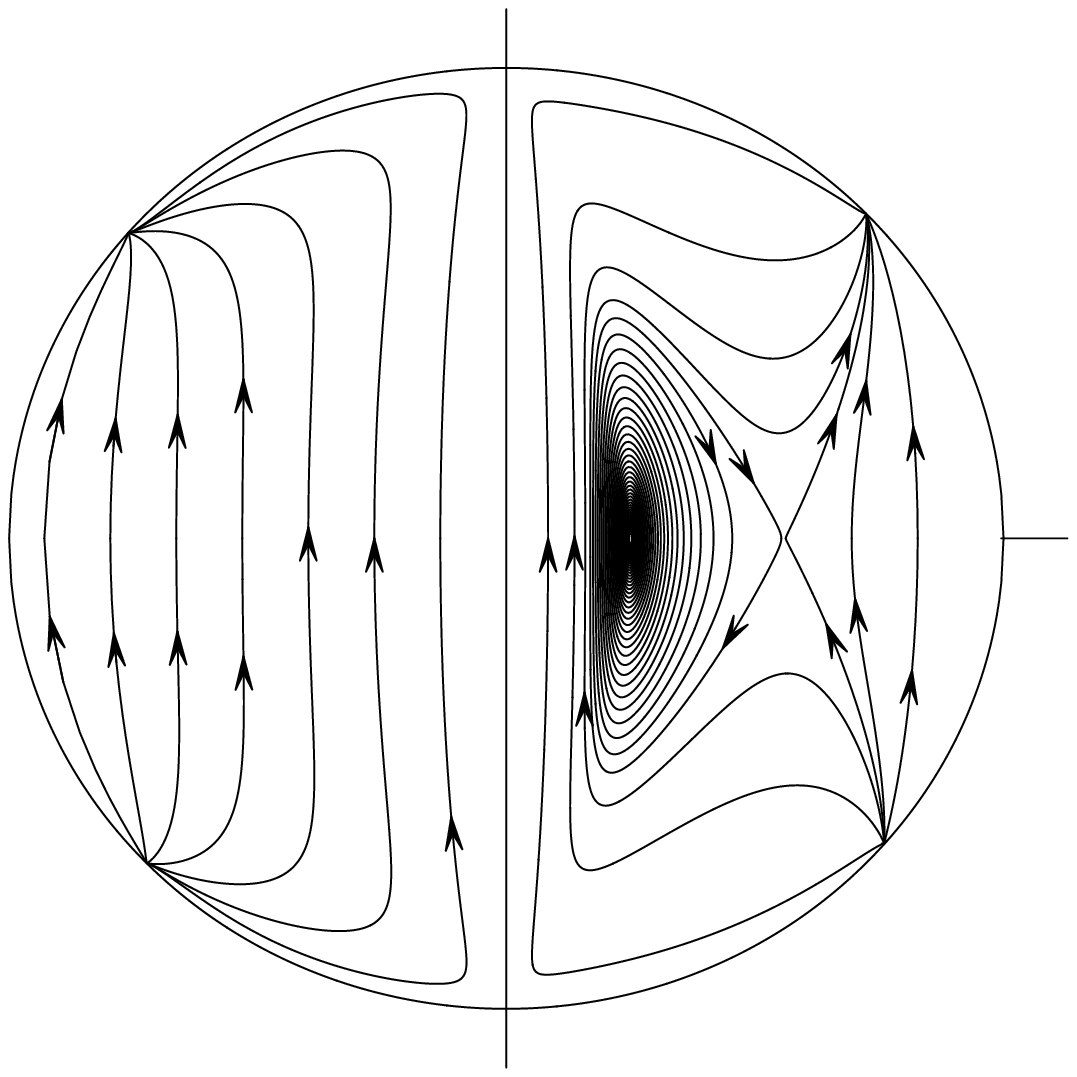}
b)\includegraphics[scale=0.75]{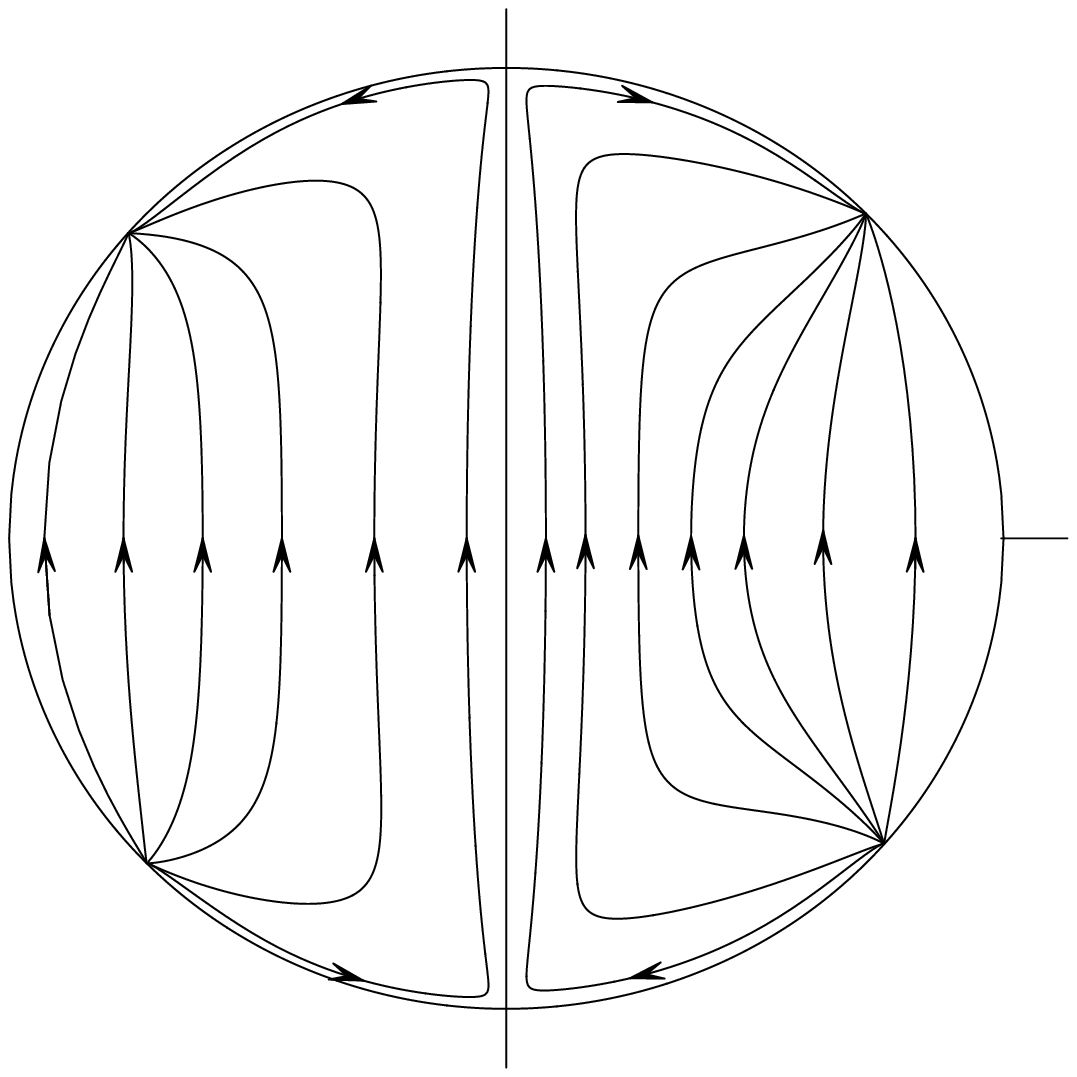}
\caption{Extended bouncing model on the Poincar{\'e} sphere. It is structurally
unstable because of the presence of the center (left figure) or 
structurally stable if this center disappear in the physically admissible domain
(right figure).}
\label{fig:10}
\end{figure}

In Figs. \ref{fig:8}, \ref{fig:9}, \ref{fig:10} we show the phase portraits of
different evolutional scenarios offered by different propositions of solving
the cosmological problem. Among different models only the $\Lambda$CDM model 
(Fig.~\ref{fig:8}) (or phantom cosmology) and bouncing cosmology in
Fig.~\ref{fig:10}a give rise to structurally stable evolutional paths.

The case of the phantom cosmology requires the additional comment. Let us 
consider the phantom cosmology with $w_{X}=4/3$ and dust matter. At the finite 
domain of the phase plane the system is described by
\begin{subequations}
\label{eq:23}
\begin{align}
\dot{x} &= y \\
\dot{y} &= - \frac{1}{2} \Omega_{\text{m},0} x^{-2} + \frac{3}{2} 
\Omega_{\text{ph},0} x^{2}
\end{align}
\end{subequations}
and possesses the first integral in the form
\[
\frac{1}{2} y^{2} + V(x) = \frac{1}{2} \Omega_{k,0}
\]
where
\[
V(x) = - \frac{1}{2} \Omega_{\text{m},0} x^{-1} - \frac{1}{2} 
\Omega_{\text{ph},0} x^{3}.
\]
To investigate the behaviour of trajectories at infinity (i.e., on the circle 
at infinity $x^{2} + y^{2} = \infty^{2}$) it is useful to introduce the 
projective coordinates on the plane. There are two maps which cover the circle 
at infinity. Let us consider one of them 
\[
(x,y) \mapsto (v,w)\colon v=\frac{1}{y}, w=\frac{x}{y}
\]
Then the circle at infinity is covered by $y=0$, $-\infty < w < +\infty$ 
(for full investigation of the system the second map $(x,y) \mapsto (z,u) 
\colon z=\frac{1}{x}, u=\frac{y}{x}$ should be studied and both coordinate 
systems are equivalent if $v \ne 0$ and $u \ne 0$). 

In the coordinates $(v,w)$ system (\ref{eq:23}) assumes the form 
\begin{subequations}
\label{eq:24}
\begin{align}
\frac{dv}{d\eta} &= \frac{1}{2} \left( \Omega_{\text{m},0} v^{5} 
- \Omega_{\text{ph},0} w^{4} v \right) \\
\frac{dw}{d\eta} &= \frac{1}{2} \Omega_{\text{m},0} v^{4}w - \frac{3}{2} w^{5} 
+ vw^{2}
\end{align}
\end{subequations}
where $dt/d\eta = vw^{2}$. Hence there is one (double) degenerated critical 
point situated at $x=0$, $y=\infty$. 

The na{\"{\i}}ve thinking about the system gives rise to the supposition that 
the system is structurally unstable because of the existence of degenerated 
critical point in infinity. The objections are the following. All solutions 
of dynamical system $x(t,x_{0})$ can be divided on two categories -- singular 
and non-singular. The former are represented by critical points in the phase 
space. The latter are visualized by trajectories joining them.  While the 
trajectories at the finite domain of the phase space represent the physical 
evolution of the evolution, the trajectories at infinity are added to the 
model by the Poincar{\'e} sphere construction and they cannot represent 
the physical solution. 

If we go along the trajectories in the future then at some moment of time, say 
$t=t_{\text{final}}$ become to tangent to the circle at infinity. This state 
is called the big-rip singularity. On the circle at infinity we can find 
a trajectory joining the big-rip singularity with the degenerate critical 
point $(x,y)=(0,+\infty)$. Note that this trajectory has no physical 
interpretation. 

In the definition of structural stability itself it appears the assumption that 
the boundary of the domain at which the system is considered  is a ``cycle 
without contact'', i.e. there is a simple smooth curve $C$ which does not 
intersect the boundary. Although this assumption bounds a class of systems it 
makes this sense of notion of structural stability to be simpler 
\cite{Bautin:1976}. 

The parameterization of time, applied for establishing the qualitative  
equivalence of phantom model with the $\Lambda$CDM model, is not of course 
a diffeomorphism. However, the phantom system in the new parameterization 
can be treated as the model of physical reality. In this case the big-rip 
singularity is reached for infinity time due to reparametrization of time 
(see Fig.~\ref{fig:9}). In this case there is no non-physical trajectory 
lying on the circle in the infinity. This transformation prolongs incomplete 
trajectories to infinity and the big-rip singularity is a global attractor. 
From the physical point of view this parameterization of the phantom cosmology 
seems to be more adequate. Therefore, we claim that the phantom cosmology is 
structural stable. However, we must remember that the existence of topological 
equivalence of trajectories of the phantom and $\Lambda$CDM models does not 
mean their physical equivalence. It is manifested by the non-diffeomorphic 
reparameterization of time. It is an example that the scientific modelling 
we can use both the fragile and structurally stable models. But some of them 
seems to be more adequate of description of physical processes. 

The cosmological models with dynamics as presented in Fig.~\ref{fig:10}a 
describe for example models in which instead of the initial singularity there 
is a characteristic bouncing phase. There are different situations which 
realize this type of evolution: the cosmological models with spinning fluid
\cite{Szydlowski:2003nv} or the metric affine gravity (MAG) inspired model
\cite{Krawiec:2005jj,Puetzfeld:2004sw}. All trajectories in Fig.~\ref{fig:10}a
represent bouncing models.

The qualitative behaviour of extended bouncing models (see Table~\ref{tab:1}) 
with the cosmological constant is illustrated in Fig.~\ref{fig:10}b. Because 
of the presence of an additional center on the phase portrait the models lost 
the property of structural stability which is the attribute of the models in 
Fig.~\ref{fig:10}a. Therefore, they are non-generic in the multiverse. 
An analogous type of evolution appeared as a phenomenological implication of 
discreteness in Loop Quantum Cosmology \cite{Singh:2005xg}. In general, if we 
consider a squeezing bounce phase in an evolutional scenario, we obtain 
a fragile model which is structurally unstable.

Recently Ashtekhar (Loops'05 Conference presentation) suggested that quantum
gravity (geometry) can serve as a bridge between vast space time regions which
are classically unrelated, i.e. in Loop Quantum Gravity, singularity is a
transitional phenomenon. Therefore, the resolution of the singularity problem
of general relativity is replaced by approach to singularity with
a bounce generated by quantum effects.

From our point of view such types of evolutional scenarios are not typical in 
the space of all evolutional paths on the plane.

Moreover, brane world models allow for a transient acceleration of the Universe
which is preceded and followed by matter domination epoch (deceleration epoch).
They admit so-called ``quiescent'' cosmological singularities
\cite{Shtanov:2002ek}, at which the density, pressure and Hubble parameter 
remain finite, while all invariants of the Riemann tensor diverge to infinity 
within a finite interval of cosmic time. From our consideration such
models are exceptional, i.e. while time-like extra dimension can avoid
cosmological singularity by bounce this proposition is not generic in the
multiverse.

The Sobolev metric introduced in the multiverse of dark energy models can be 
used to measure how far different cosmological model with dark energy are to 
the canonical $\Lambda$CDM model. For this aim let us consider a different dark
energy models with dust matter and dark energy. We also for simplicity of
presentation assume for all models have the same value of $\Omega_{\text{m},0}$ 
parameters which can be obtained, for example, from independent extragalactic
measurements. Then, the distance between any two cosmological model, say model 
``1'' and model ``2'' is
\[
d(1,2)=\max_{x \in C}\{|V_{1x}-V_{2x}|,|V_{1xx}-V_{2xx}|\},
\]
where we assumed the same value of the $H_{0}$ parameter measured at the present
epoch for all cosmological models which we compare, $V_{1}$ and $V_{2}$ and 
their derivatives are only the parts of the potentials without the matter term.

Because $(V_{1}-V_{2})_{x}$ measures the difference of slopes at any $x$, the
distance $d$ in general becomes a function of $x$. However, for fitting or
constraining a model's parameters we use actual observational data obtained at
$x=1$ epoch. Therefore, a closed region $C$ on which we compare predictions of
different theoretical models can be chosen such that $C=\{1\}\times\{y \colon 
0 < y \le \sqrt{-2V(1)}, V(1)=-1/2\}$. Of course, it is a closed domain. From 
the definition of the metric $d$, two models are close if the slopes of
tangents at the present epoch are close. This metric can also be expressed in
dimensionless parameters, then we simply obtain the deceleration parameter 
instead of slopes. Also, instead of $C^{1}$ metric $C^{r}$ metric can be 
defined. Then $C^{r}$ metric measures the distance between any two models more 
precisely by comparing their additional acceleration indicators like jerk, 
snap, crackle or higher derivatives of potential functions. In general, the 
metric is a function of the model's parameters. Therefore with estimating the 
values of parameters, one can immediately determine the distance between any 
two models -- elements of an ensemble of dark energy models. Because at present 
we can only measure the second derivative of scale factor as an acceleration 
indicator, $C^{1}$ metric seems to be sufficient to order variety of dark 
energy models for answering how far the model is from the $\Lambda$CDM one and 
any other. In this way we obtain a ranking of cosmological models from the 
point of view of closeness to the concordance $\Lambda$CDM model. Different 
models belong to the open ball at the center of which the $\Lambda$CDM model is 
located. One can show the existence of the following inclusion relation
\cite{Szydlowski:2006gp,Szydlowski:2006se}
\[
S_{\Lambda \text{CDM}} \subset S^{(w_{X}\text{fitted})}_{\text{phantom}} 
\subset S^{w_{X}=-4/3}_{\text{phantom}} \subset S_{\text{Cardassian}} \cdots
\]
It is interesting that the above ordering is correlated with one obtained from 
the Bayesian information criterion \cite{Szydlowski:2006pz,Szydlowski:2006ay}.

\section{Conclusions}
\label{sec:5}

The main goal of this paper was to investigate the structure of the space of 
all FRW models which offer the possibility of explaining SN Ia data. To this 
aim we presented a unified language of dynamical systems of the Newtonian type 
in which the potential function determines all properties of the system. We 
defined the space of all dynamical systems, called the multiverse, of 
accelerating models. This space can be naturally equipped with the structure 
of Banach space which measures the distance between two models. This metric 
can be used to measure how far different model are from the concordance 
$\Lambda$CDM model.

The complexity of different models can be defined in terms of the potential
function. This function determines the domains of the configuration space in
which the Universe accelerates or decelerates. The concept of structural
stability was used to distinguish generic models in the multiverse. Following 
Peixoto's theorem we called them typical in the multiverse because they form 
open and dense subsets. The structurally unstable models are exceptional and 
form sets of zero measure in the multiverse. Our main result is that the 
structural stability property uniquely determines the shape of the potential. 
It is shown that genericity favours inverted single-well shape of the potential 
function.

On the other hand, this function could be reconstructed (modulo curvature) from
distant type Ia supernovae data and the obtained function is equivalent to the
potential function for the $\Lambda$CDM model. The evolutional scenario in which
the acceleration epoch is proceeded by deceleration is uniquely distinguished.
Therefore, while different theoretically allowed evolutional scenarios have been
proposed and formulated in terms of the potential function, the simplicity of
the evolutional scenario is the best guide to our Universe -- the inverted 
single-well potential function is preferred. In other words, our Universe shows
evidence of complexity and at the same time great simplicity which allows us to
probe its properties with the help of simple models.

In explanation and understanding of the observational data we use the models 
(an idea of cosmological models). However, between the physical subjects and 
theoretical notions there is a 1--1 correspondence preserving some relations 
(isomorphism). 

If we require the property of structural stability of the model we implicitly
apply what McMullin called cosmogonic indifference principle 
\cite{McMullin:1993}. While the anthropic like principles concentrated on
explaining some properties of the Universe by the specially chosen model
parameter, initial conditions or laws of physics, the indifference type of
explanation concentrate on searching upon very generic initial conditions and
laws of physics which act to produce the special configuration (see
\cite[p. 18]{Stoeger:2004sn}). Stoeger, Ellis and Kirchner argue that indifference
principle is more interesting from the physical point of view relative to
anthropic type of explanation which is more useful in philosophy. The authors
claim that the problem of choosing between two principles: special or generic 
model has a rather philosophical (or epistemological) character to which the 
interpretation of our results is strictly related.

Therefore if we concentrate on searching for a very generic class of dark 
energy models or modification of the FRW equation that produce the special 
configuration we now enjoy -- an accelerating Universe, then the postulate of 
structural stability naturally gives rise to such a situation.

A natural question is whether the inverted single-well potential is favoured
over other more complex models with double, triple etc. accelerating phases.
This can be addressed by using the Bayesian information criteria (BIC) of model
selection. Our analysis confirms that there is no strong reason for inclusion
of extra complexity (more accelerating epochs) and the model with a single 
acceleration epoch is favoured over the others by supernovae data
\cite{Godlowski:2005tw,Szydlowski:2005qb}.

Different conclusions can be made from our analysis but the answer to the 
question put in the title seems to be especially tempting. We are living in 
an accelerating Universe because simplicity is the best guide to our Universe. 
Moreover, because our Universe is typical (generic) and its model can be 
discovered by using the approximation method starting from a simple (possibly 
naive) model -- the $\Lambda$CDM (or phantoms).

\acknowledgments{
The paper was supported by Marie Curie Host Fellowship MTKD-CT-2004-517186 
(COCOS) and was substantially developed during staying in University of 
Paris 13. The author is very grateful to dr Adam Krawiec for discussion, 
comments, and help with preparing the final version of the paper. I would like 
to thank prof. R. Kerner and prof. J. Madore for discussion on the structural 
stability in the cosmological context. I also thank prof. J.-M. Alimi and his 
group for useful comments during the colloquium in the Meudon Observatory.}

\end{document}